\RequirePackage{fix-cm}
\documentclass{svjour3} 
\smartqed 
\usepackage{comment}
\usepackage{color}
\usepackage[titletoc,title]{appendix}
\usepackage{amsmath}
\usepackage{amssymb}
\usepackage[authoryear]{natbib}
\usepackage{graphicx}
\usepackage{array}
\setcitestyle{notesep={; },round,aysep={},yysep={;}}
\usepackage[colorlinks,linkcolor=blue,anchorcolor=red,citecolor=green]{hyperref}
\usepackage{multirow}
 \usepackage{mathptmx} 
\def\DJ{\mbox{\raise0.3ex\hbox{-}\kern-0.4em D}}

\begin{document}

\title{Secular resonances between bodies on close orbits II:}


\subtitle{prograde and retrograde orbits for irregular satellites}

\titlerunning{Secular resonances between bodies on close orbits II} 

\author{Daohai Li \and
 Apostolos A. Christou 
}


\institute{D. Li \at
 Armagh Observatory, College Hill, Armagh, BT61 9DG, Northern Ireland, UK \\
 School of Mathematics and Physics, Queen's University Belfast, University Road, Belfast, BT7 1NN, Northern Ireland, UK\\
 Tel.: +44 (0)28 3752 2928\\
 Fax: +44 (0)28 3752 7174\\
 \email{lidaohai@gmail.com; dli@arm.ac.uk} 
 \and
 A. A. Christou \at
 Armagh Observatory, College Hill, Armagh BT61 9DG, Northern Ireland, UK 
}

\date{Received: date / Accepted: date}

\maketitle

\begin{abstract}
In extending the analysis of the four secular resonances between close orbits in \citet{Li2016} (Paper I), we generalise the semianalytical model so that it applies to both prograde and retrograde orbits with a one to one map between the resonances in the two regimes. We propose the general form of the critical angle to be a linear combination of apsidal and nodal differences between the two orbits $ b_1 \Delta \varpi + b_2 \Delta \Omega $, forming a collection of secular resonances in which the ones studied in Paper I are among the strongest. Test of the model in the orbital vicinity of massive satellites with physical and orbital parameters similar to those of the irregular satellites Himalia at Jupiter and Phoebe at Saturn shows that $> 20\%$ and $> 40\%$ of phase space is affected by these resonances, respectively. The survivability of the resonances is confirmed using numerical integration of the full Newtonian equations of motion. We observe that the lowest order resonances with $b_1+|b_2|\le 3$ persist, while even higher order resonances, up to $b_1+|b_2|\ge 7$, survive. Depending on the mass, between $10\%-60\%$ of the integrated test particles are captured in these secular resonances, in agreement with the phase space analysis in the semianalytical model.

\end{abstract}

\section{Introduction}
\label{sec-intro}

Resonances are common in celestial mechanics. They may excite the action-like orbital elements, for example, the eccentricity and inclination, and possibly destabilise the motion.

Mean motion resonances (MMRs) may occur when the mean motions of two or more orbits are near-commensurable. MMRs are prevalent in the solar system and among exoplanets \citep[see, e.g.][]{Gallardo2006,Fabrycky2014}. For the HD 73526 planetary system, \citet{Gayon2008} found that a counter-revolving configuration could provide more stability while remaining consistent with observations; the system's stability was related to retrograde MMRs and especially to apsidal precession. \citet{Gayon2009} analytically investigated the retrograde 2:1 MMR by introducing a new set of canonical variables. In \citet{Morais2012}, the authors developed analytical models to compare the properties of a given MMR between retrograde and  prograde orbits in the circular planar restricted three-body problem; the cause of MMR-related instability was addressed. \citet{Morais2013a} identified several retrograde asteroids in MMRs with Jupiter or Saturn as ``the first examples of Solar system objects in retrograde resonance''. In \citet{Morais2013}, a dedicated semianalytical approach for the retrograde 1:1 MMR was described and its properties were presented.

In addition to MMRs, the properties of the retrograde evection resonance are distinct from those of the prograde case. In developing their lunar theory, \citet{Brouwer1961a} identified evection as an important periodic perturbation. The evection perturbation is proportional to $\cos (2 \lambda_\odot-2\varpi)$, where $\lambda_\odot$ is the mean longitude of the Sun and $\varpi$ the lunar longitude of pericentre. When $\dot \lambda_\odot \sim \dot \varpi$, the lunar eccentricity $e$ could be pumped up; this is generally the evection phenomenon. \citet{Touma1998} showed that, the Moon, in its outward migration, could have passed through the evection resonance, possibly exciting the eccentricity.

In the context of irregular satellites, \citet{Nesvorny2003} reported that evection could be a key factor in determining orbital stability and specifically in causing their asymmetric distribution with respect to an inclination of $90^\circ$. More recently, \citet{Yokoyama2008} analytically studied the evection dynamics, showing how the resonant structure was different for prograde and retrograde orbits; e.g., the libration centres. \citet{Frouard2010} extended this result to higher order expansions of the disturbing potential, obtaining new asymmetries within each regime.

\citet{Christou2005} identified a type of nodal secular resonance locking the nodes of two close orbits in a family of prograde irregular satellites. In \citet{Li2016} (Paper I hereafter), we used a semianalytical method to model the resonance and in combination with $N$-body simulation, three more types were observed. Motivated by the fact that MMR and evection resonances exist for both prograde and retrograde and that their properties can be distinct, we want to explore whether the above mentioned resonances can persist when the orbits are retrograde and if so, whether the properties change.

The paper is organised as follows: in Sect. \ref{sec-n-retrograde}, we construct a ``retrograde'' group of satellites by flipping the orbits investigated in Paper I. Through numerical integrations of the full equations of motion we show that retrograde resonances such as those studied in Paper I exist. We then introduce a semianalytical model (Sect. \ref{sec-semi}) to show that the retrograde and prograde resonances are identical, to demonstrate the existence of higher order resonances and to study their properties. In Sect. \ref{sec-semi-apply} we test the semianalytical model with massive satellites with physical and orbital parameters similar to those of the irregular satellites Himalia at Jupiter and Phoebe at Saturn. Sect. \ref{sec-n-high} is dedicated to identifying prograde and retrograde resonances in the numerical simulations and to comparing their properties. In the Sect. \ref{sec-statistics}, we discuss the likelihood of these resonances and their mean lifetime. Finally, we summarise our results and discuss some of the implications in Sect. \ref{sec-conclusion}.
\section{Retrograde resonances in $N$-body simulations}
\label{sec-n-retrograde}
In \citet{Li2016} (Paper I) we studied secular resonances within a restricted four-body problem comprised of Jupiter, the Sun, a massive Himalia and a test particle serving as a fictitious group member, all revolving around Jupiter. In the numerical simulations, 1000 particles were integrated for 100 Myr. Between 10 and 30 particles were found to librate in each of the four resonances.

In searching for similar resonances in the retrograde case, we first employ $N$-body simulations. For the purpose of making comparisons between prograde and retrograde resonances, it is ideal to keep all parameters unchanged except the inclination. Thus we simply flip the satellite orbits used in Paper I -- the new orbital inclination is the supplement of the original value, i.e., $i\approx 150^\circ$. In this way, the orbits of the massive satellite and 1000 test particles in Paper I are reversed to create retrograde orbits. We integrate these orbits for 100 Myr with the $N$-body package MERCURY \citep{Chambers1999,Hahn2005} using the general Bulirsch-Stoer algorithm with a tolerance of $10^{-12}$. For simplicity, this set of simulations will be referred to as case i2 while the one studied in Paper I will be referred to as case i1. For the full set of physical and orbital parameters used in the simulations, see Table \ref{tab-param-case}.

As mentioned in Sect. \ref{sec-intro}, the resonances found in the prograde case may also exist for retrograde orbits but their properties may be different. Using an automated method to detect episodes of libration may fail to find all or most resonant episodes. We resort to visually inspecting the evolution of the corresponding angles. For this reason we then need to decide which angles may librate as inferred from Paper I.

A commonly used set of variables to describe the orientation of an orbit in space is the argument of pericentre $g$ and longitude of ascending node $h$ \citep[in Delaunay notation and see, e.g.,][]{Murray1999}; we refer to those as the {\it basic} set of variables. Another set of angles is $\varpi=h+g$ and $\Omega=h$ for prograde orbits; we call these the {\it derived} variable set; for retrograde orbits, the former becomes $\varpi=h-g$ \citep{Whipple1993}. In the following, we will adopt the derived variable set.

The four angles presented in Paper I, when expressed in the derived variables, are $\Delta\Omega$, $\Delta\varpi$, $\Delta\varpi-\Delta\Omega$ and $\Delta\varpi+\Delta\Omega$. We expect that the resonant angles in the prograde and retrograde case may either remain (i) in the same form in derived variables, in which case we do not need to do anything, or (ii) identical in terms of the basic variable set; in the latter case we need to convert the prograde derived variables to basic variables first and then transform these basic variables into retrograde derived variables. For example, the resonant angle $\Delta \varpi +\Delta \Omega$, through route (i), preserves its form in the retrograde situation, but through route (ii) it becomes $\Delta \varpi +\Delta \Omega \rightarrow \Delta (h+g) +\Delta h \rightarrow -\Delta (h-g) +3\Delta h \rightarrow -\Delta \varpi+3\Delta \Omega$, a different angle altogether. Similarly, the other three prograde angles, through those two routes, may be mapped to more than three angles. Actually, we have a total of six angles that may be librating. They are $\Delta \Omega$, $\Delta \varpi$, $\Delta\Omega-\Delta\varpi$, $\Delta\varpi+\Delta\Omega$, $\Delta\varpi-2\Delta\Omega$ and $\Delta\varpi-3\Delta\Omega$.

\begin{figure}[h]
\begin{center}
	\includegraphics[width=\textwidth]{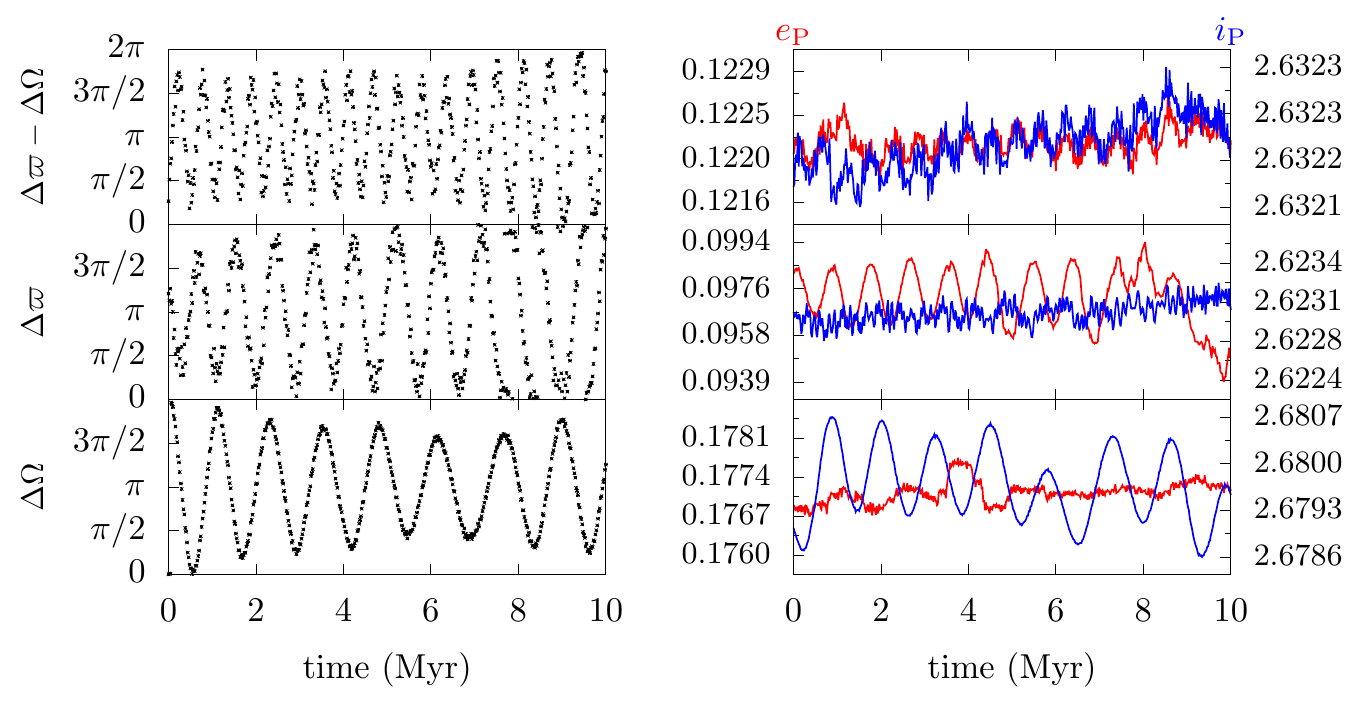}
\caption{Examples of librating particles in the $N$-body simulations for the angles $\Delta\varpi-\Delta\Omega$ (top), $\Delta\varpi$ (middle) and $\Delta\Omega$ (bottom) in case i2. Left column: the critical angle; right column: the eccentricity $e$ (red, left $y$-axis) and inclination $i$ (blue, right $y$-axis); panels in the same row correspond to the same particle; $x$-axis: time in Myr. All angles are measured in rad
}
\label{fig-retro-n-6-1}
\end{center}
\end{figure}

\begin{figure}[h]
\begin{center}
	\includegraphics[width=\textwidth]{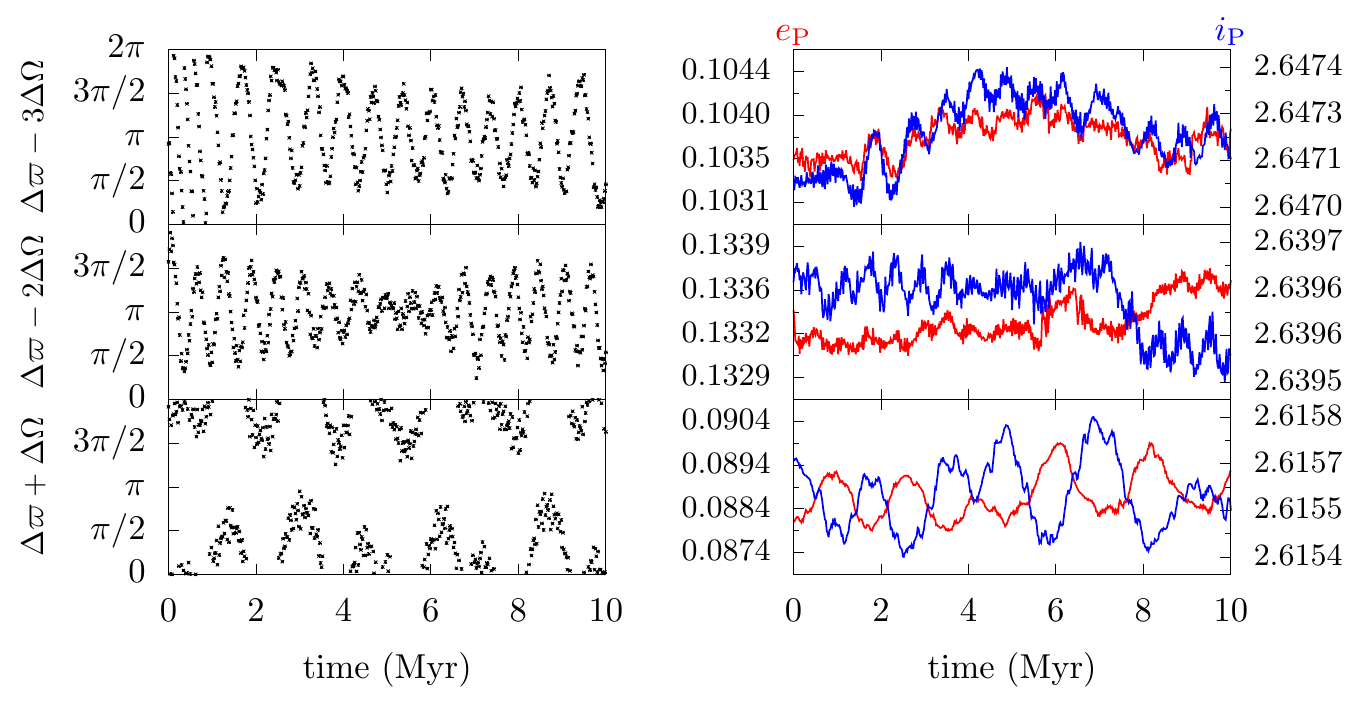}
\caption{As Fig. \ref{fig-retro-n-6-1} but for the angles $\Delta \varpi- 3\Delta \Omega$, $\Delta \varpi- 2\Delta \Omega$ and $\Delta \varpi+\Delta \Omega$}
\label{fig-retro-n-6-2}
\end{center}
\end{figure}

By plotting the evolution of these angles for all 1000 retrograde particles in the $N$-body simulations and through visual inspection, we find that all six angles can librate. Examples of resonant episodes are shown in Figs. \ref{fig-retro-n-6-1} and \ref{fig-retro-n-6-2}.

The resonance involving the angle $\Delta \Omega$, the nodal resonance for retrograde orbits, is shown in the bottom panels of Fig. \ref{fig-retro-n-6-1}. Comparing it with the prograde case discussed in Paper I, we see that the two are qualitatively identical; the libration centre is $\pi$ and $i$ experiences periodic oscillations whereas there are only low amplitude, short-period variations in $e$. Also, $\Delta\Omega$ and $i$ evolve in phase: when $i$ is near maximum, $\Delta\Omega$ is increasing and vice versa. An example of what we refer to as the ``apsidal'' resonance is shown in the middle panels of the same figure. As with prograde motion, $e$ oscillates in phase with the critical angle $\Delta \varpi$. Note that the libration centre is at $0$ for the prograde case whereas here it is $\pi$. Also, the retrograde apsidal resonance appears more well-behaved than the prograde one in the sense that, here $\Delta \varpi$ and $e$ are clearly phase-correlated while for the prograde case, such a link was not apparent (cf. fig.~15 of Paper I). The top panels of Fig. \ref{fig-retro-n-6-1} show the evolution of a librating particle for the critical angle $\Delta\varpi-\Delta \Omega$. Here, $e$ and $i$ are correlated whereas, for prograde orbits, they are anti-correlated. As in the case of the retrograde Kozai mechanism \citep{Carruba2002}, we can explain the difference using the conservation of the vertical angular momentum: if we use the Delaunay angle-action set $(h,H;g,G)$, the resonant angle for both prograde and retrograde orbits is $\Delta g$, meaning that the momentum conjugate to the fast angle $\Delta h$, $H\propto \sqrt{1-e^2}\cos i$ is conserved. The change in the slope of $\cos i$ depending on whether $i$ is smaller or larger than $\pi/2$ determines whether $e$ \& $i$ are positively or negatively correlated.

In Fig. \ref{fig-retro-n-6-2} we show resonant episodes for the remaining three angles $\Delta \varpi- 3\Delta \Omega$, $\Delta \varpi- 2\Delta \Omega$ and $\Delta \varpi+\Delta \Omega$, from top to bottom. For the resonance of $\Delta \varpi+\Delta \Omega$, the angle librates about $0$, the same as for prograde orbits but $e$ and $i$ are anti-correlated, contrary to the prograde case. This can be explained using the similar argument as made above for the angle $\Delta\varpi-\Delta \Omega$. In addition, the two new resonances in $\Delta \varpi- 2\Delta \Omega$ and $\Delta \varpi- 3\Delta \Omega$ both have libration centres around $\pi$. Meanwhile, $e$ and $i$ oscillate and they are correlated.

We have shown that the angles corresponding to the four prograde resonances identified in Paper I may be converted to six angles in the retrograde case depending on how we perform the conversion. We then find through $N$-body simulations that all these angles can librate. This leads us to wonder whether the two new angles $\Delta \varpi- 2\Delta \Omega$ and $\Delta \varpi- 3\Delta \Omega$ can librate for prograde orbits as well. Revisiting our $N$-body simulations in case i1 in Paper I, we find that examples of libration of these two angles do exist. 

The process of generating the six angles from the original four can be applied again to these two angles. For a prograde angle of the general form $b_1 \Delta \varpi + b_2 \Delta \Omega$ (where $b_1$ and $b_2$ are integer coefficients and we require $b_1 \ge 0$; $b_1$ and $b_2$ to be mutually prime) or, equivalently, $(b_1+b_2)\Delta h+b_1\Delta g$. Its counterpart for retrograde orbits can be either $b_1 \Delta \varpi + b_2 \Delta \Omega$ (through route (i)) or $-b_1\Delta\varpi+(2b_1+b_2)\Delta\Omega \sim b_1\Delta\varpi-(2b_1+b_2)\Delta\Omega$ (through route (ii)). The prograde equivalent of the second retrograde angle is different from the original angle $b_1 \Delta \varpi + b_2 \Delta \Omega$ (when $b_1>0$ and through route (i)). Repeating this procedure may give rise to more and more angles. Otherwise, an efficient way to look for resonant angles is to construct a semianalytical model as in Paper I and reveal the inherent link between the resonant angles in prograde and retrograde orbits.

\section{A unified semianalytical model for prograde and retrograde orbits}
\label{sec-semi}
A semianalytical model valid for prograde orbits was presented in Paper I. There, we split the restricted four-body problem consisting of a planet, the Sun, a massive satellite and a test particle into two separate restricted three-body problems. The planet-Sun-satellite (either the massive one or the test particle) sub-problem was modelled using the Kozai-Lidov formalism \citep[e.g.,][]{Lidov1962,Kozai1962,Naoz2013}. The solar potential was expanded in terms of the ratio of semimajor axis of the satellite to that of the Sun $a/a_\odot$. This Kozai potential caused the satellite's orbit to precess on timescales $\sim 10^2$ yr. The second sub-problem, comprised of the planet, massive satellite and the test particle was modelled using secular coorbital theory \citep{Henon1986,Luciani1995,Namouni1999}. Assuming the eccentricity and inclination of the satellites to be small, the secular coorbital potential was expressed in the relative elements between the particle and  the massive satellite. Then the two restricted three-body sub-problems were combined into a system of coupled nonlinear equations. Noting the libration timescales of these resonances ($\sim 1$ Myr), the much faster precession induced by the Kozai potential was removed. This resulted in a four-dimensional dynamical system containing only the angle differences $\Delta \varpi$ and $\Delta \Omega$, the eccentricity $e_\mathrm{P}$ and inclination $i_\mathrm{P}$ of the particle. By integrating this system numerically, we reproduced the three resonances in $\Delta \Omega$, $\Delta \varpi-\Delta \Omega$, and $\Delta \varpi+\Delta \Omega$ (See Paper I for details) We expect that a similar model can be constructed for retrograde orbits and we show this in the following.

We begin by noting that the inclinations of retrograde irregular satellites usually have $i \gtrsim 140^\circ$ \citep[e.g., ][]{Carruba2002} where the coorbital theory originating from \citet{Henon1986} does not apply. The other component of the model, the Kozai potential, only involves an expansion in terms of a ratio between semimajor axes and is therefore suitable for retrograde orbits \citep[for example,][]{Naoz2013}. \footnote{The particular Kozai Hamiltonian used in Paper I included expansion in inclination; see eq. (12) of that paper.}

To obtain an appropriate form of this coorbital potential for the retrograde case one can expand the inclination around $\pi$ as done by \citet{Morais2013} for retrograde MMRs, then follow the derivation of \citet{Henon1986,Luciani1995}. However, this might be a cumbersome procedure. Instead, we introduce a new reference system: the ``flipped'' frame \citep[e.g.,][]{Saha1993}, in which the original coorbital potential is valid. To arrive at this frame, which we denote by the triad $x'$-$y'$-$z'$, we rotate the original frame $x$-$y$-$z$ about the $x$-axis by $\pi$ (see Fig. \ref{fig-flip-frame} in the appendix). In the original frame, the $x$-$y$ plane is the orbital plane of the Sun and $z$ is along the direction of the solar angular momentum; in the flipped frame, the Sun is still in the $x'$-$y'$ plane, but it is now retrograde with an inclination of $\pi$. The principal advantage of this transformation is that the originally retrograde satellites become prograde in the new frame, suiting the coorbital theory. Thus we only need to obtain the Kozai potential in the flipped frame to complete our model. Following the detailed derivation in \citet{Valtonen2006}, we verify that the Kozai potential (see Paper I) is valid in the flipped frame and retains its form as in the original frame. Hence, we can describe the evolution of the retrograde satellite in the flipped frame using exactly the same semianalytical model as in Paper I.

In the appendix, we show that the orbital elements of a satellite: the semimajor axis $a$, eccentricity $e$, inclination $i$, argument of pericentre $g$, longitude of ascending node $h$ and true anomaly $f$, in the two frames, obey the following relationships:
\begin{equation}
\label{eq-original-flipped-1}
\begin{aligned}
a'&=&a,\,\qquad f'&=& f;
\\
e'&=&e,\,\,\,\,\qquad g' &=& g+\pi;
\\
i'&=& \pi-i,\qquad h'&=&\pi-h.
\end{aligned}
\end{equation}
A primed notation means the the quantity is measured in the flipped frame throughout the paper; those not primed represent quantities in the original frame.

We consider the angular differences between two members of a retrograde irregular satellite group
\begin{equation}
\Delta h'=-\Delta h,\qquad \Delta g' =\Delta g.
\end{equation}
As mentioned in Sect. \ref{sec-n-retrograde}, we will be using derived variables defined as $\Omega=h$ and $\varpi=h-g$ for retrograde orbits in the original frame; in the flipped frame, the prograde orbit has $\Omega'=h'$ and $\varpi'=h'+g'$. Thus for an arbitrary combination of $\Delta \varpi$ and $\Delta \Omega$
\begin{equation}
\label{eq-original-flipped-2}
\begin{aligned}
b_1\Delta \varpi+b_2\Delta \Omega&=&b_1(\Delta h-\Delta g)+b_2\Delta h
\\
&=&b_1(-\Delta h'-\Delta g')-b_2\Delta h'
\\
&=&-(b_1\Delta \varpi'+b_2\Delta \Omega').
\end{aligned}
\end{equation}
This means that there exists a one-to-one correspondence between prograde and retrograde resonant angles.


\section{Application of the model}
\label{sec-semi-apply}

To demonstrate the dynamics, we test the model with different parameters. One goal of the paper is to extend the results of Paper I to retrograde orbits, so consideration of a similar system is desirable. Thus we test the model on (i) a massive satellite with the same parameters as in Paper I except that we make the orbits retrograde using transformation \eqref{eq-original-flipped-1}. In addition, to show the dependence of the results on the satellite's mass, $e$ and $i$, we test our model on (ii) another massive satellite with about 10 times the relative massive (compared to the host planet) and smaller $i'$. These choices will be referred to as (parameter) sets i and ii; c.f. Table \ref{tab-param-case}. We note that the physical and orbital parameters of saturnian irregular satellite Phoebe have been used for the massive satellite in set ii. Though no apparent family is currently found around it \citep[e.g.,][]{Gladman2001}, we believe this parametric choice is appropriate to explore the dependence of the resonant dynamics on different parameters. Throughout the paper, the term ``set'' is used in semianalytical model while ``case'' is used in $N$-body simulations.

\subsection{Examples of resonant motion}
\label{sec-individual}
In Paper I, we introduced the concept of the Surface of Equal Precession Rate (SEPR), defined as the surface in $(a,e,i)$ space where all points on it have the same rate of change of an angle as the massive satellite, i.e., ${\mathrm{d}(b_1\Delta \varpi+b_2\Delta \Omega) / \mathrm{d}t}=0$ and showed that a resonance involving the angle $b_1 \Delta\varpi+b_2 \Delta\Omega$ could only exist near its corresponding SEPR. Therefore, to systematically search for additional resonances we have randomly placed particles on each SEPR with $b_1+|b_2|\le 10$, a total of 64 separate angles. We integrate these particles for 8 Myr in the semianalytical model and visually check if the angles can librate. We find that this is the case for all the angles with the exception of $\Delta\varpi$.\footnote{The semianalytical model is unable to reproduce this particular resonance; see Paper I.} The libration centres of these resonances are all 0 or $\pi$ and the  libration timescales are all $\sim 1$ Myr. For ease of reference, we refer to the sum $b_1+|b_2|$ as the ``order'' of the critical angle.

Each critical angle, when in or near libration, precesses much more slowly than $\Delta \varpi$ and/or $\Delta \Omega$, as long as the particle is not close to the common intersection of all the SEPRs (where both ${\mathrm{d}\Delta \varpi / \mathrm{d}t}=0$ and ${\mathrm{d}\Delta \Omega / \mathrm{d}t}=0$). We can then introduce a transformation to isolate the resonant motion. The angular momentum conjugate to the fast angle in the transformed system is proportional to
\begin{equation}
\label{eq-e-i-correlation}
b_2 \bar G' -b_1\bar H'\propto b_2 e'^2 -b_1 i'^2=b_2 e^2 -b_1 (\pi-i)^2
\end{equation}
and it should be conserved on the resonant timescale. Hence, for a resonance involving the angle $b_1\Delta \varpi+b_2\Delta \Omega$, when $b_2$ is positive, the eccentricity and inclination are anti-correlated for prograde orbits and correlated for retrograde orbits; the opposite is true when $b_2$ is negative. We offer this as an explanation for the behaviour of $e$ and $i$ in Sect. \ref{sec-n-retrograde}.

\begin{figure}[h]
\begin{center}
	\includegraphics[width=\textwidth]{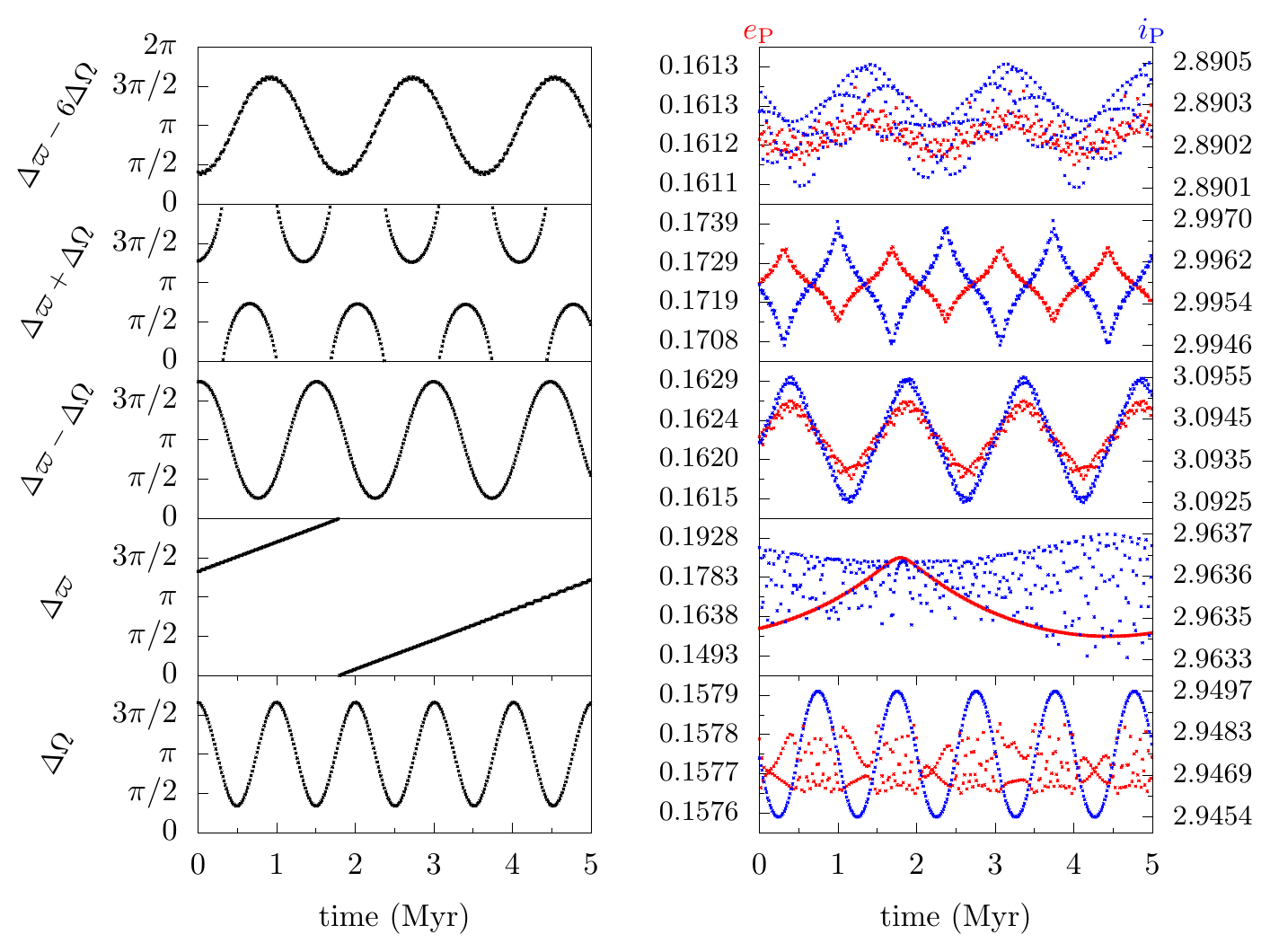}
\caption{As Fig. \ref{fig-retro-n-6-1} but for the angles $\Delta \varpi- 6\Delta \Omega$, $\Delta \varpi+\Delta \Omega$, $\Delta \varpi-\Delta \Omega$, $\Delta \varpi$ and $\Delta \Omega$ and for parameter set i in semianalytical model}
\label{fig-semi-new-1}
\end{center}
\end{figure}

In Fig. \ref{fig-semi-new-1}, we show the resonant examples of the 5 angles $\Delta \varpi- 6\Delta \Omega$, $\Delta \varpi+ \Delta \Omega$, $\Delta \varpi- \Delta \Omega$, $\Delta \varpi$ and $\Delta \Omega$ for parameter set i. The last four were presented in Paper I for prograde orbits. Here we use these to demonstrate how the properties of the resonances differ in terms of the libration centre and the behaviour of $e$ and/or $i$. The evolution of $e$ and $i$ is consistent with the analysis of Eq. \eqref{eq-e-i-correlation}. 
Libration of $\Delta \varpi- 6\Delta \Omega$, a 7th order argument according to our definition, is shown as a representative of a high order resonance. We note that the amplitudes of $e$ and $i$ are smaller and that the amplitude of short period variations is of the same order as that induced by the resonance.

\begin{figure}[h]
\begin{center}
	\includegraphics[width=\textwidth]{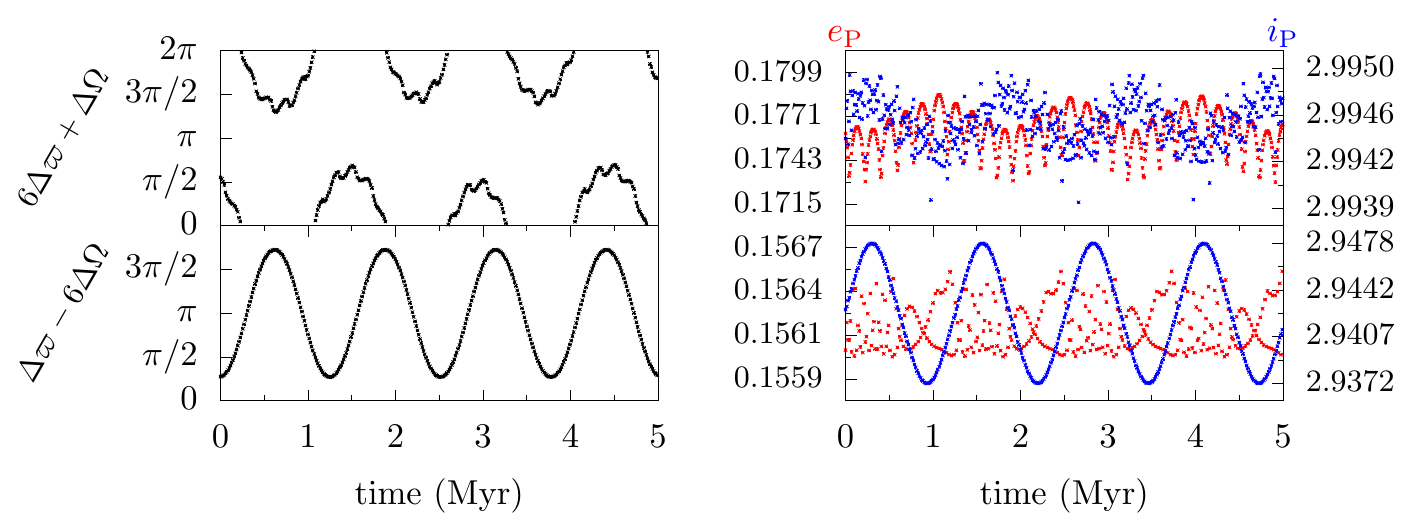}
\caption{As Fig. \ref{fig-semi-new-1} but for parameter set ii and for the angles $6\Delta \varpi +\Delta\Omega$ and $\Delta\Omega$}
\label{fig-semi-new-2}
\end{center}
\end{figure}

In Fig. \ref{fig-semi-new-2}, we present two resonant examples for parameter set ii. The nodal resonance (bottom panels) behaves in a similar fashion as in parameter set i. Note here, the amplitude of the inclination is larger, probably due to the larger mass in set ii. We will discuss the influence of mass latter. A resonance that involves the 7th order argument $6\Delta \varpi+ \Delta \Omega$ is shown in the top panels. Again, we observe smaller amplitude variations similar in magnitude to the short period variations.

For these individual resonances, we can use a procedure similar to that introduced in Paper I to eliminate the fast angle (cf. discussion on Eq. \eqref{eq-e-i-correlation}) to arrive at a two-dimensional system that encapsulates the resonant dynamics. Since we observe only one libration centre for each critical angle, we expect that the phase space structure of a resonance with libration centre $\pi$ is similar to that of the resonances of $\Delta\Omega$ and $\Delta \varpi-\Delta \Omega$ (see figs. 8 and 10 of Paper I) and one with libration centre $0$ is analogous to that of $\Delta\varpi+\Delta\Omega$ (fig. 11 of Paper I). 



\subsection{Resonant strength}
\label{sec-strength}

\begin{figure}[h]
\begin{center}
	\includegraphics[width=0.8\textwidth]{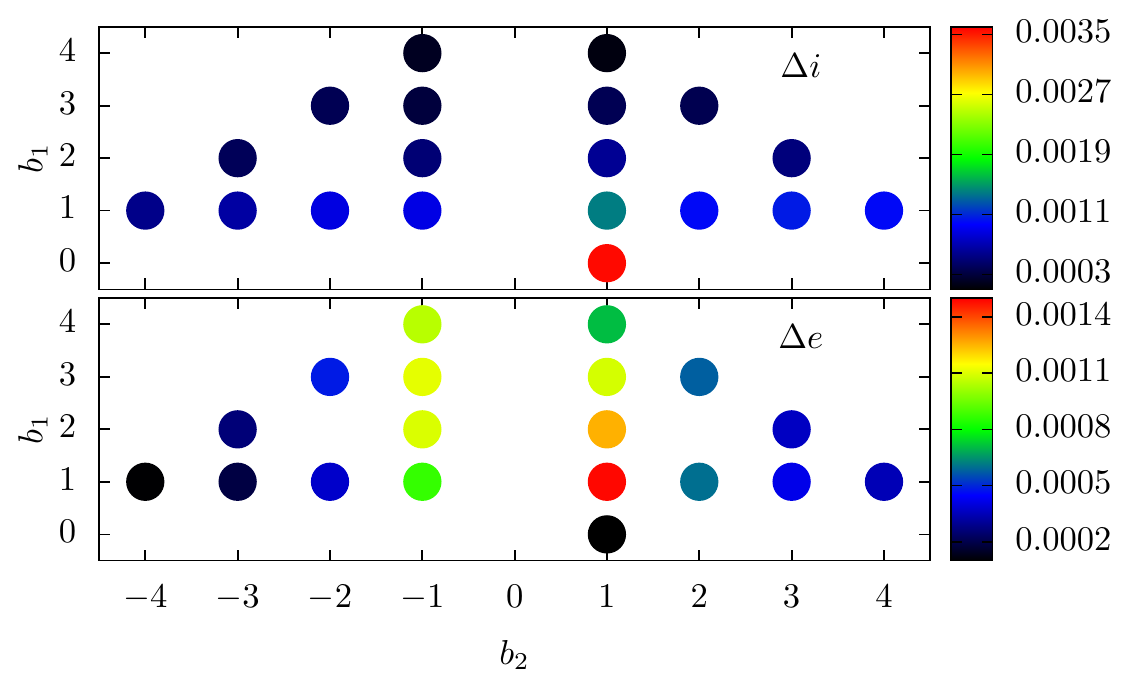}
\caption{The average strength of a resonance for the angle $b_1\Delta \varpi+b_2\Delta \Omega$ for parameter set i. From top to bottom, the colour coding shows the mean values of $\Delta i=i_\mathrm{max}-i_\mathrm{min}$ and $\Delta e= e_\mathrm{max}-e_\mathrm{min}$ for each resonance. Warmer colours represent larger amplitudes. $x$-axis: integer coefficient of $\Delta \Omega$, $b_2$; $y$-axis: integer coefficient of $\Delta \varpi$, $b_1$}
\label{fig-amp}
\end{center}
\end{figure}

In Paper I, we used the amplitude of the action during libration as a proxy for resonance strength. It is of interest here to rank the different resonances in terms of this strength. For this purpose, we have placed 40 test particles on each of the SEPRs of resonances up to order 5 and integrated them also for 8 Myr. The mean resonant amplitudes $\Delta e $ and $\Delta i $ are shown in Fig. \ref{fig-amp} for parameter set i. A general observation is that, the higher the order of the critical angle, the smaller the amplitude, in both $\Delta e$ and $\Delta i$. This suggests that the perturbation corresponding to the angle $b_1\Delta \varpi+b_2\Delta \Omega$ may be proportional to the actions raised to the exponent $b_1+|b_2|$, for instance
\begin{equation}
\label{eq-order}
(e')^{2b_1} (i')^{2|b_2|},
\end{equation}
in line with a traditional expansion of the secular perturbing potential \citep[see e.g., chapter 6 of][]{Murray1999}. This seems to support our use of $b_1+|b_2|$ as the order of the resonance. But we emphasise that this is an empirical result to be confirmed by a formal series development of the disturbing potential.

Recall that we have shown that for libration of the angle $b_1\Delta \varpi+b_2\Delta \Omega$, the quantity $b_2 e'^2 -b_1 i'^2$ is conserved (cf. Eq. \eqref{eq-e-i-correlation}). Consequently, we have $|b_2|\Delta( e'^2) = b_1\Delta ( i'^2)$. Noting the transformation \eqref{eq-original-flipped-1}, we have 
\begin{equation}
\label{eq-delta}
|b_2| e \Delta e =b_1  i' \Delta i'=b_1 i' \Delta i.
\end{equation}
Thus the amplitudes $\Delta e$ and $\Delta i$ should be inversely proportional to $b_2$ and $b_1$, respectively. We suppose Eqs.~\eqref{eq-order} and \eqref{eq-delta} contribute to the amplitudes. This is observed in Fig.~\ref{fig-amp}. For example, for two resonances of the same order, the one with a larger $b_1$ (thus a smaller $|b_2|$) has larger $\Delta e$ and smaller $\Delta i$, and vice versa.

Interestingly,  a resonance involving the angle $b_1\Delta \varpi+|b_2|\Delta \Omega$ shows a higher amplitude in both $e$ and $i$ than $b_1\Delta \varpi-|b_2|\Delta \Omega$. In addition to this, we observe that the nodal resonance has the largest amplitude in $i$, about three times of that of the second strongest, $\Delta \varpi+\Delta \Omega$; the latter angle shows the largest $\Delta e$.

As described in Sect. \ref{sec-individual}, the amplitude of the variation in $i$ induced by the nodal resonance for parameter set ii is larger than that for set i. Note that the coorbital potential is proportional to the mass ratio of the massive satellite to the central planet, $m_{\mathrm{massive\,\, satellite}}/m_{\mathrm{planet}}$ \citep{Luciani1995}. This value is $\sim 2.2\times 10^{-9}$ for set i while for set ii it is $\sim 1.5\times 10^{-8}$, about an order of magnitude larger. Thus we expect $\Delta e$ and $\Delta i$ to be larger for the latter set.

\begin{figure}[h]
\begin{center}
	\includegraphics[width=0.8\textwidth]{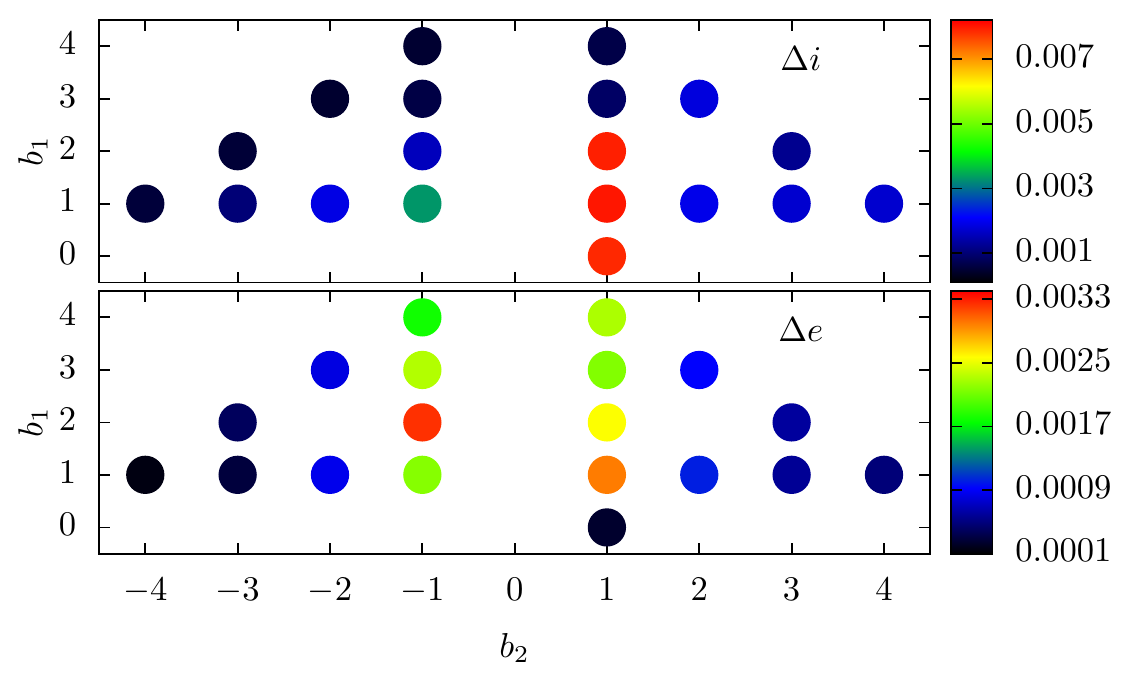}
\caption{As Fig. \ref{fig-amp} but for parameter set ii}
\label{fig-amp-Phoebe}
\end{center}
\end{figure}

Mean amplitudes in $e$ and $i$ for resonances of up to order 5 for parameter set ii are shown in Fig. \ref{fig-amp-Phoebe}. Amplitudes of resonances here are about twice as large as those for set i. Specifically, here the resonance with largest $\Delta e$ is $2\Delta\varpi-\Delta \Omega$; in addition, resonances of $\Delta\varpi+\Delta \Omega$ and $2\Delta\varpi+\Delta \Omega$ have similar amplitude in $i$ to the nodal resonance. Thus, the relation between argument order and resonance strength appears to be more complex than speculated through \eqref{eq-order} and \eqref{eq-delta}.

\subsection{Resonant region}
\label{sec-region}
A resonance has a finite width in terms of the variation in the resonant action. This width demarcates a volume occupied by the resonance in the form of a sheet in phase space.

By approximating MMR dynamics as a pendulum, the resonance width can be analytically derived \citep[e.g.][]{Nesvorny1998,Murray1999}. However, here we are not yet at a stage where we can isolate terms associated with a particular critical angle in the secular potential \citep[originally from][see eq. (17) of Paper I for the specific form we have used]{Luciani1995}.
For secular resonances in the asteroid belt, \citet{Milani1990,Milani1994} prescribed a maximum deviation from exact zero precession of a resonant angle as an indicator of resonance width. However, the arguments that correspond to the resonances studied here span 5 orders, making it inconvenient to provide each resonance with a width measured in terms of the maximum allowable precession rate. Alternatively, since we already have the mean amplitudes $\Delta e$ and $\Delta i$ of a resonance from the semianalytical model in the last subsection, we use those as empirical estimates of the resonance width.

\begin{figure}[h]
\begin{center}
	\includegraphics[width=0.9\textwidth]{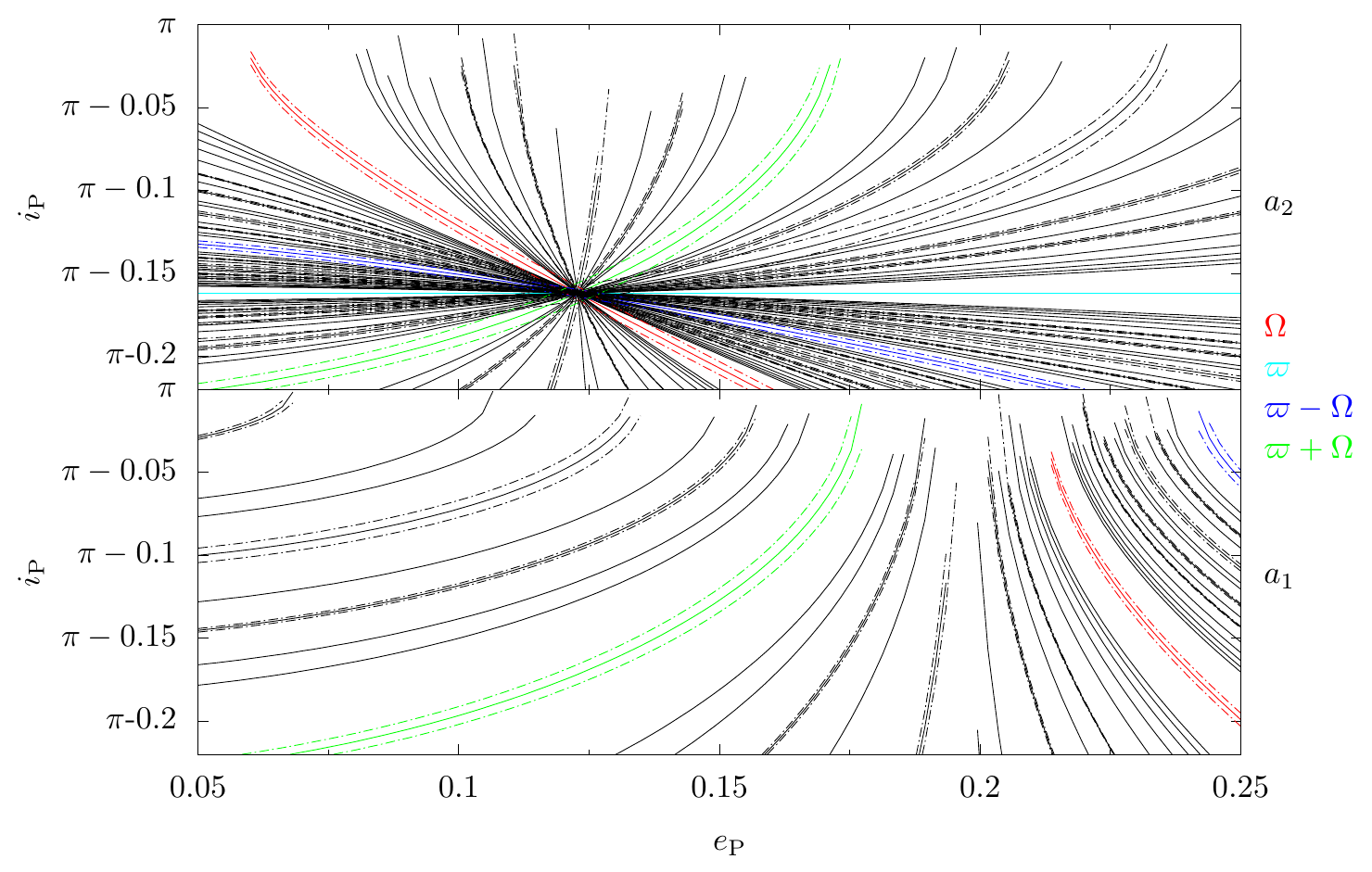}
\caption{Intersections of the SEPRs and the resonant regions of the angles $b_1 \varpi+b_2 \Omega$ for parameter set ii with the planes $a=a_1=0.98 a_\mathrm{massive\,\, satellite}$ (lower panel) and $a=a_2=1.02 a_\mathrm{massive\,\, satellite}$ (upper panel). $x$-axis: eccentricity; $y$-axis: inclination. Solid curves: plane intersections of the SEPRs with $b_1+|b_2|\le10$; dash-dotted curves: resonant regions for resonances with $b_1+|b_2| \le 5$; see text for details}
\label{fig-SEPR}
\end{center}
\end{figure}

For a retrograde resonance involving the angle $b_1\Delta \varpi+b_2\Delta \Omega$, the variation amplitudes in $e$ and $i$ are $\Delta e=e_\mathrm{max}-e_\mathrm{min}$ and $\Delta i=i_\mathrm{max}-i_\mathrm{min}$, respectively. From the conservation of the quantity \eqref{eq-e-i-correlation}, when $ b_2>0$ the evolution of $e$ and $i$ is anti-correlated. Then any point $(e_0,i_0)$ on the SEPR will be largely evolving along the line segment linking the two points $(e_0-\Delta e /2, i_0+\Delta i /2)$ and $(e_0+\Delta e /2, i_0-\Delta i /2)$. Similarly, when $b_2<0$, the same holds for the two points $(e_0-\Delta e /2, i_0-\Delta i /2)$ and $(e_0+\Delta e /2, i_0+\Delta i /2)$. In this way, each SEPR can be mapped to two parallel surfaces, forming a sheet of finite width between them. The enclosed volume is the corresponding resonant region. A resonance with larger $\Delta e$ and $\Delta i$ has a larger resonant region associated with it. In addition, the resonant region also depends on the strength in a more subtle way; for example, a resonant region cannot be large if the vectors $(0,\Delta e, \Delta i)$ or $(0,\Delta e, -\Delta i)$ are nearly perpendicular to the norm of the SEPR in $(a,e,i)$ space. Resonant regions for parameter set ii are shown in Fig.~\ref{fig-SEPR} for angles up to order 5; the mean resonant amplitudes are from Fig. \ref{fig-amp-Phoebe}. In addition, we show SEPRs for resonances up to order 10 in the same figure. The two panels correspond to the cuts $a=a_1$ and $a=a_2$, similar to how \citet{Milani1994} illustrated the location of secular resonances in the asteroid belt. We choose $a_1=0.98 a_\mathrm{massive\,\, satellite}$ and $a_2=1.02 a_\mathrm{massive\,\, satellite}$, as in fig. 20 of Paper I, where we showed the SEPRs of the four resonances known at that time for set i (prograde orbits). Comparing the two sets of SEPRs for prograde and retrograde orbits, we see essentially a vertically flipped version of what was presented in Paper I. This follows Eq. \eqref{eq-original-flipped-1}.

It is clear that $(a,e,i)$ space is densely populated by these resonances. We also observe that the resonant regions for different resonances can overlap.

While libration is only possible inside its resonant region, being inside a resonant region does not guarantee libration. As pointed out by \citet{Carruba2005,Carruba2009a}, the only criterion to decide whether a critical angle is librating or not is to plot its time variation. Knowing a resonant region, we can estimate its size relative to the entire phase space -- this is the likelihood that a particle randomly placed near a massive satellite is influenced by this resonance. It follows that the result depends on our definition of ``entire'', which in turn depends on the orbital distribution of the particles.

Irregular satellite groups are thought to be remnants of collisional evolution \citep[e.g.,][]{Nesvorny2003}. Upon fragmentation, the relative velocity $\delta v$ among different members caused dispersion in $a$, $e$, $i$. Following \citet{Beauge2007}, $\delta v$ can be expressed as
\begin{equation}
\delta v^2={\mu\over \langle a\rangle (1-\langle e\rangle^2)}\left[{\left(1-\langle e\rangle ^2\right)^2\over4\langle e\rangle^2} \left({\delta a \over \langle a\rangle}\right)^2+ {1\over 2} (\delta e)^2 + 2 (\delta \sin i)^2\right],
\end{equation}
where $\langle a\rangle$ and $\langle e\rangle$ are the mean semimajor axis and eccentricity of the family; the orbital element notation preceded by the $\delta$ symbol represents the difference from the mean. Typically, $\delta v \lesssim 100$ m/s for a collisional group \citep[e.g.][]{Michel2001,Michel2004}. However, for the Himalia group, it is observed that $\delta v >300$ m/s \citep[][see also footnote 7 of Paper I]{Beauge2007}, some of which could have resulted from internecine gravitational scattering \citep{Christou2005}. For our purposes we generate two groups, each containing $10^6$ randomly located particles, one with $\delta v< 100$ m/s and another with $\delta v< 300$ m/s, representing a compact and a loose group respectively. Then we count how many particles are inside a resonant region. This procedure is applied to massive satellites from parameter sets i and ii. We emphasise that we are not trying to model hypothetical groups or families associated with actual satellites. Instead, we only use the above criterion to delimit the $(a,e,i)$ space near a massive satellite and what is of interest is the dynamical property of the space.

Table \ref{tab-region-size} shows that, for a massive satellite from parameter set ii, the resonance involving the angle $2\Delta \varpi+\Delta\Omega$ affects more than $10\%$ of the particles in a compact group near it but only $4 \%$ of a loose group. The second strongest resonance is $\Delta \varpi+\Delta\Omega$, affecting $8.8\%$ and $2.8\%$ of the compact and loose groups, respectively. In total, we expect that such an irregular satellite would affect more than $40\%$ of its vicinity with resonances up to order 5 $b_1+|b_2|\le 5$.

For a satellite from parameter set i, the strongest resonance corresponds to the angle $\Delta \varpi+\Delta \Omega$, affecting $4\%$ of a compact group and $1.3\%$ of a loose group. The angles $2\Delta \varpi+\Delta\Omega$ and $\Delta\Omega$ rank second and third in terms of the relative size of the resonant region. Overall, $23.6\%$ and $8.4\%$ of the compact and loose group are affected, respectively. 

Note that, if we sum up the fractions of $(a,e,i)$ space that each resonant region takes up in a compact group hosted by a satellite with parameter set ii, we obtain $51.0 \%$ (resonances of order 5 are not shown in Table \ref{tab-region-size}). However, the total fraction of particles affected by the resonances is $42.1\%$. We speculate that the difference, $9\%$ consists of those affected by more than one resonance, as noted for Fig. \ref{fig-SEPR}.

Further analysis of the table will be presented in Sect. \ref{sec-statistics} with direct comparison with $N$-body simulation results.

\begin{table}[h]
\caption{Relative size of a resonant region as a fraction of $(a,e,i)$ space enclosed by a particular value of $\Delta v$ for parameter sets i and ii of a resonance $b_1\Delta \varpi+b_2\Delta \Omega$. First column shows the resonant angle. The second and third columns correspond to a compact group with $\delta v<100$ m/s: left is for set ii and right is for set i. The last two columns are for a loose group with $\delta v<300$ m/s: left is for set ii and right is for set i. Only individual resonances with $b_1+|b_2| \le 4$ are shown. The last row shows the total relative size of all resonant regions with $b_1 +|b_2| \le 5$.}
\label{tab-region-size}
\begin{center}
\begin{tabular}{ *5{>{\centering\arraybackslash}m{0.16\textwidth}} @{}m{0pt}@{}}
\hline
\hline
\multirow{2}{*}{resonant angle} & \multicolumn{2}{c}{compact group ($\Delta v<$100 m/s)} &\multicolumn{2}{c}{loose group ($\Delta v<$300 m/s)}\\
{} & {set ii}& {set i}& {set ii}& {set i} \\
 \hline
$\Delta\Omega$& 2.4&	 2.4	& 0.8	& 0.8\\
$\Delta \varpi$& -& -& -& -\\
$\Delta \varpi-\Delta\Omega$& 4.3&	 2.1	& 1.4	& 0.7\\
$\Delta \varpi+\Delta\Omega$& 8.8&	 4.0	& 2.8	& 1.3\\
$\Delta \varpi-2\Delta\Omega$& 2.0&	 1.4	& 0.7	& 0.5\\
$\Delta \varpi+2\Delta\Omega$& 1.6&	 1.1	& 0.4	& 0.4\\
$2\Delta \varpi-\Delta\Omega$& 3.5&	 1.6	& 1.1	& 0.5\\
$2\Delta \varpi+\Delta\Omega$& 11.7&	 2.8	& 3.9	& 0.9\\
$\Delta \varpi-3\Delta\Omega$& 0.8&	 0.9	& 0.3	& 0.3\\
$\Delta \varpi+3\Delta\Omega$& 0.7&	 0.5	& 0.2	& 0.2\\
$3\Delta \varpi-\Delta\Omega$& 1.8&	 1.1	& 0.6	& 0.4\\
$3\Delta \varpi+\Delta\Omega$& 3.0&	 1.8	& 1.0	& 0.6\\
all $b_1 +|b_2| \le 5$&42.1 &23.6&15.6&8.4\\
\hline
\end{tabular}
\end{center}
\end{table}

Out of the 63 types of resonance that we have identified in our exploration of the semianalytical model, 5 survive for both prograde and retrograde orbits in numerical simulations as shown in Sect. \ref{sec-n-retrograde}. Hence, we are left with 58 candidate critical angles that we wish to check if they can librate and how their properties are preserved when the full dynamics of the problem are accounted for.

\section{Further resonances in $N$-body simulations}
\label{sec-n-high}
As in Sect. \ref{sec-n-retrograde}, we have integrated 1000 test particles designated as case i2. In addition, we generate 1000 test particles around a satellite with parameter set ii which we refer to as case ii. Together with the 1000 prograde particles from Paper I (case i1), they form a sample of  $3\times1000$ test particles numerically integrated for 100 Myr with MERCURY \citep{Chambers1999,Hahn2005}. The physical and orbital parameters of the massive satellite in the three cases are listed in Table \ref{tab-param-case}.  We use the same method as in Paper I to detect resonant episodes, as we can predict the libration centres through the semianalytical model. In brief, our code searches for episodes of libration based on the behaviour of the critical angle -- generally we require that it oscillate around the libration centre for at least 2-3 libration cycles. Then we record the time the particle enters and leaves the resonance. Furthermore, we visually inspect all the detected resonant episodes to identify false positives. The overall false positive rate in identifying resonant episodes is $\lesssim 2\%$.

\begin{table}[h]
\caption{Physical and orbital parameters used in $N$-body simulations: cases i1, i2 and ii and in semianalytical model: parameter sets i and ii. In cases i1, i2 and set i, we use the same parameter values except for the inclination; those in case ii and set ii are identical.}
\label{tab-param-case}
\begin{center}
\begin{tabular}{c c c}
\hline
\hline
 & case i1/case i2/set i & case ii/set ii\\ 
 \hline
   & Himalia &Phoebe\\
 $m_{\mathrm{massive\,\, satellite}}/m_{\mathrm{planet}}$ & $ 2.2\times 10^{-9}$&$ 1.5\times 10^{-8}$\\
 $a$ (AU)&$0.076$&$0.087$ \\
$e$& $0.16$&$0.16$ \\
$i$& $28.6^\circ$/$151.4^\circ$/$170.5^\circ$&$175.0^\circ$ \\
\hline
\end{tabular}
\end{center}
\end{table}


\subsection{ Resonances in case i2}
\label{sec-n-high-retrograde}
\begin{figure}[h]
\begin{center}
	\includegraphics[width=\textwidth]{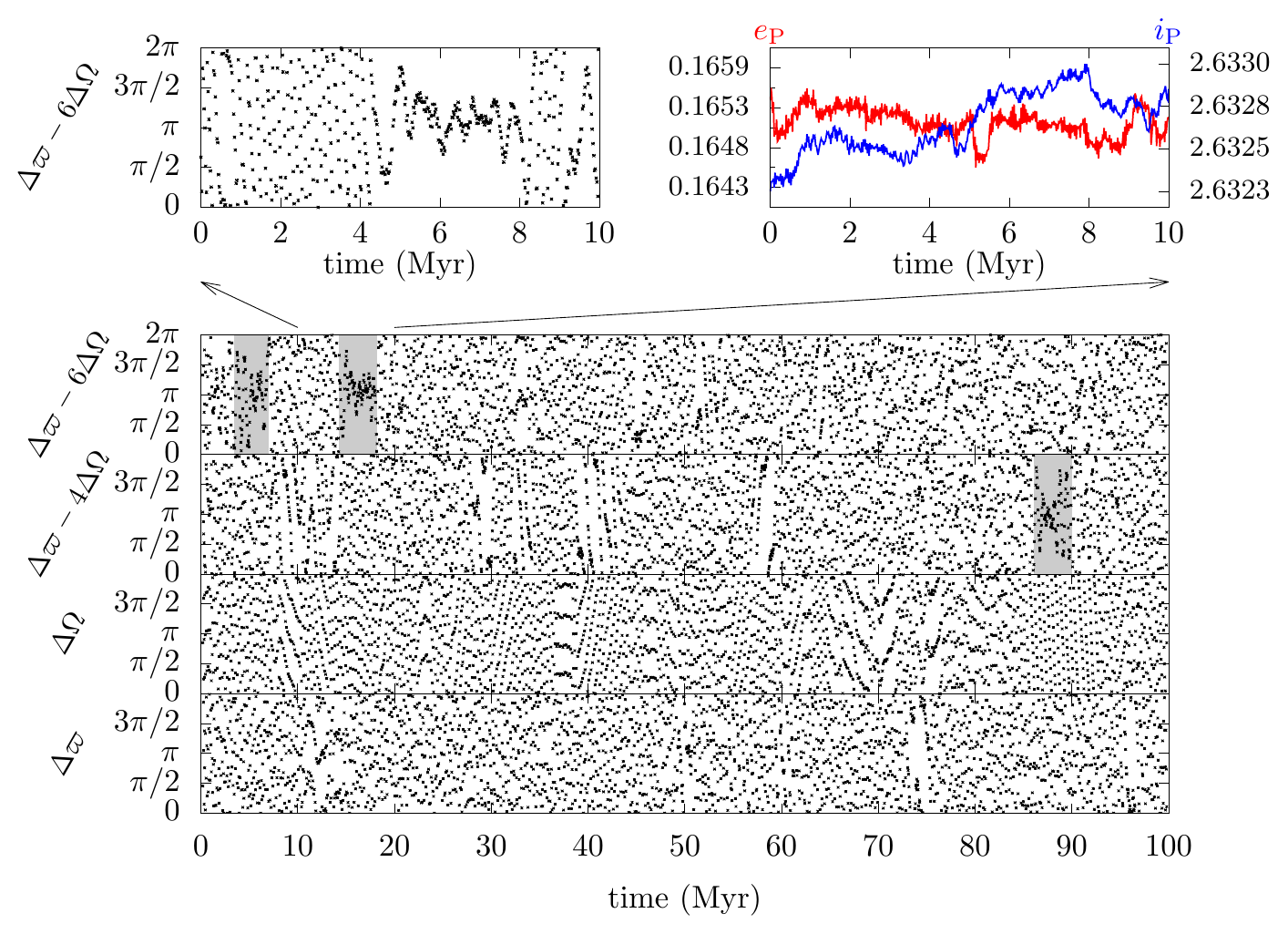}
\caption{A resonant particle found in $N$-body simulation case i2 and for the angle $\Delta\varpi -6\Delta\Omega$. Top two panels are as Fig.~\ref{fig-retro-n-6-1}. The four panels below are for the same particle, showing evolution of the angles, from top to bottom, $\Delta\varpi -6\Delta\Omega$, $\Delta\varpi -4\Delta\Omega$, $\Delta\Omega$ and $\Delta\varpi$ for the entire integration. The shaded areas mark detected libration episodes
}
\label{fig-retro-n-10-1}
\end{center}
\end{figure}

In the case  i2, in addition to the 6 confirmed types of resonance discussed in Sect. \ref{sec-n-retrograde}, we have found 10 more: $\Delta\varpi +2\Delta\Omega$, $2\Delta\varpi -\Delta\Omega$, $2\Delta\varpi +\Delta\Omega$, $\Delta\varpi+3 \Delta\Omega$, $\Delta\varpi -4\Delta\Omega$, $\Delta\varpi +4\Delta\Omega$, $2\Delta\varpi -3\Delta\Omega$, $\Delta\varpi-5 \Delta\Omega$, $\Delta\varpi -6\Delta\Omega$ and $\Delta\varpi -7\Delta\Omega$. In Fig.~\ref{fig-retro-n-10-1} we show an example of libration for the 7th order argument $\Delta\varpi -6\Delta\Omega$ as representative of a high order resonance; cf. Fig. \ref{fig-semi-new-1} for the same resonance in the semianalytical model. In the resonance shown here, the libration centre agrees with the prediction of the semianalytical model. To check if the libration centre can be different from the one in the semianalytical model, we have searched for libration around a centre that is $\pi$ away from the predicted one (e.g., if the centre is 0, we try $\pi$). None is found, indicating that the semianalytical model correctly predicts the libration centre in the full problem. However, the behaviour of the eccentricity and inclination is less regular in the simulations. This is likely due to factors omitted in the semianalytical treatment, for instance, effects related to the mean motions and higher order terms in the expansion of the secular potential. Indeed, as noted in \citet{Saha1993}, the fast variations in eccentricity and inclination are larger than the secular oscillations for the four then known retrograde irregular satellites at Jupiter. This is particularly important for the high order resonances that are relatively weak.

It is important to emphasise that the observed libration of an angle of the general form $b_1\Delta\varpi+b_2\Delta\Omega$ is independent of the evolution of the component angles $\Delta\varpi$ or $\Delta\Omega$. To highlight this, we use the libration episode in $\Delta \varpi-6\Delta\Omega$ shown in Fig. \ref{fig-retro-n-10-1}. The bottom four panels show the detection of two resonant episodes; we focus on the second one to show the evolution of the critical angle and of $e$ \& $i$. When $\Delta \varpi-6\Delta\Omega$ is librating, $\Delta \varpi$ and $\Delta\Omega$ do not. Indeed, the in-phase evolution of the resonant angle and the actions in the semianalytical model strongly suggests a true resonant phenomenon. As in \citet{Greenberg1975}, the effects of the massive satellite are enhanced by the commensurability of the frequencies of the particle and it (recall that the resonances are located near the SEPRs). We also notice that, for this specific particle, the angle $\Delta\varpi -4\Delta\Omega$ librates for a few Myr around 88 Myr into the integration, a potential indication of resonance overlap.

\subsection{Resonances in case i1}
\label{sec-n-high-prograde}

Checking the 1000 prograde particles in case i1 originally studied in Paper I, we detect 19 types of resonance: the 16 types found in the retrograde case in Sects. \ref{sec-n-retrograde} and \ref{sec-n-high-retrograde}; and 3 additional types that involve the angles $2\Delta\varpi+3\Delta\Omega$, $2\Delta\varpi-5\Delta\Omega$ and $2\Delta\varpi+5\Delta\Omega$. Therefore, 19-4=15 new types of prograde resonance are observed in the $N$-body simulations.

\subsection{Resonances in case ii}
\label{sec-phoebe-resonance}

\begin{figure}[h]
\begin{center}
	\includegraphics[width=\textwidth]{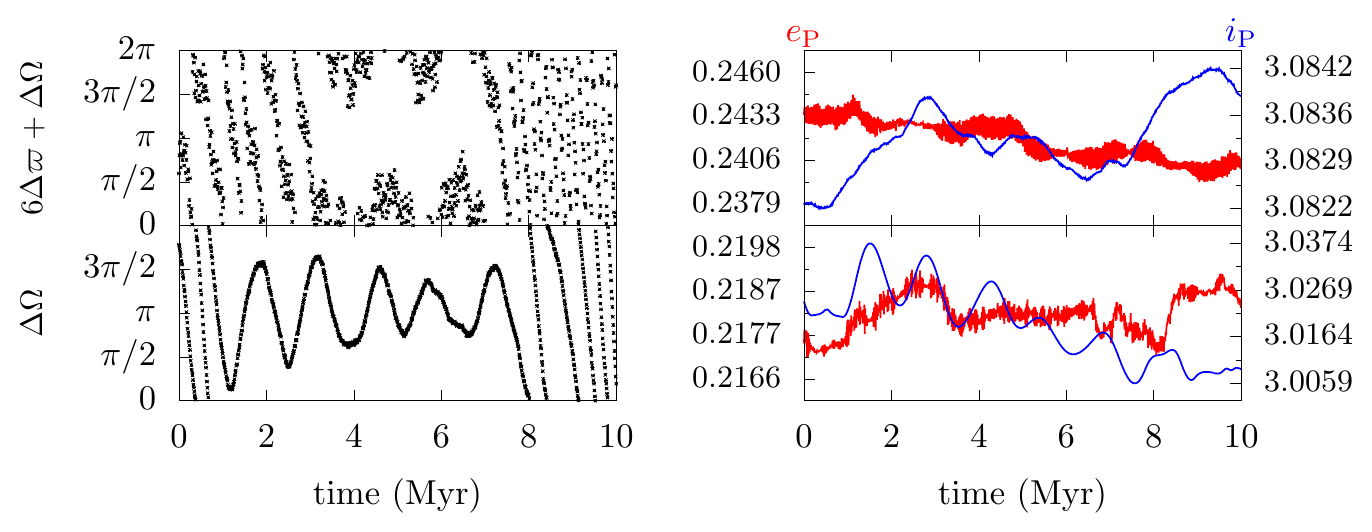}
\caption{As Fig. \ref{fig-retro-n-6-1} but for angles $6\Delta \varpi+\Delta \Omega$ and $\Delta \Omega$ and in case ii}
\label{fig-phoebe-n}
\end{center}
\end{figure}

17 types of resonance in case ii are identified, involving the angles $\Delta\Omega$, $\Delta\varpi$, $\Delta\varpi-\Delta\Omega$, $\Delta\varpi+\Delta\Omega$, $\Delta\varpi-2\Delta\Omega$, $\Delta\varpi+2\Delta\Omega$, $2\Delta\varpi-\Delta\Omega$, $2\Delta\varpi+\Delta\Omega$, $3\Delta\varpi-\Delta\Omega$, $3\Delta\varpi+\Delta\Omega$, $3\Delta\varpi+2\Delta\Omega$, $4\Delta\varpi-\Delta\Omega$, $4\Delta\varpi+\Delta\Omega$, $5\Delta\varpi+\Delta\Omega$, $5\Delta\varpi+2\Delta\Omega$, $6\Delta\varpi+\Delta\Omega$ and $7\Delta\varpi+2\Delta\Omega$. In Fig. \ref{fig-phoebe-n} we show two examples, the nodal resonance and one involving the 7th order argument $6\Delta\varpi+\Delta\Omega$.

It is worthwhile to note that in case ii the coefficient of the detected resonance $b_1$ can be large while $b_2$ is always small whereas in cases i1 and i2, the opposite is true. This may be related to how the strength of the resonance depends on $e$ \& $i$ of the moons (Eq.~\eqref{eq-order}). In case ii, $e\sim 0.16$ and $i'=\pi-i\sim 0.08$ while for cases i1 and i2 $e\sim 0.16$ and $i\sim 0.5$ (or $i'\sim 0.5$). Hence, in cases i1 and i2 the coefficient \eqref{eq-order} depends more sensitively on $b_1$ (since $e$ is smaller) but more sensitively on $b_2$ in case ii (as $i'$ is smaller). Thus in case ii, a resonance with a large $b_2$ may be too weak to occur in the simulations.

Even though in the examples shown here, we have deliberately chosen episodes where the resonance is relatively long-lived, individual episodes are shorter than the 10 Myr interval shown, it is natural to ask the questions of how often and how long particles are trapped in resonances. We study this in the following.
\section{Statistics of $N$-body simulations}
\label{sec-statistics}
Having collected a sample of numerous resonant episodes among the $3\times 1000$ test particles, we can quantify their frequency. For the four angles  $\Delta \Omega$, $\Delta \varpi$, $\Delta \varpi-\Delta \Omega$ and $\Delta \varpi+\Delta \Omega$, those were calculated for case i1 in Paper I. With the additional integrations and the appearance of high order resonances, we can now compare (i) prograde with retrograde resonances and (ii) low order with high order resonances; we discuss the results for cases i1 and i2 in Sect.~\ref{sec-statistics-Himalia} and for case ii in Sect. \ref{sec-phoebe-statistics}.

\subsection{Statistics of resonances in cases i1 and i2}
\label{sec-statistics-Himalia}
\begin{figure}[h]
\begin{center}
	\includegraphics[width=0.9\textwidth]{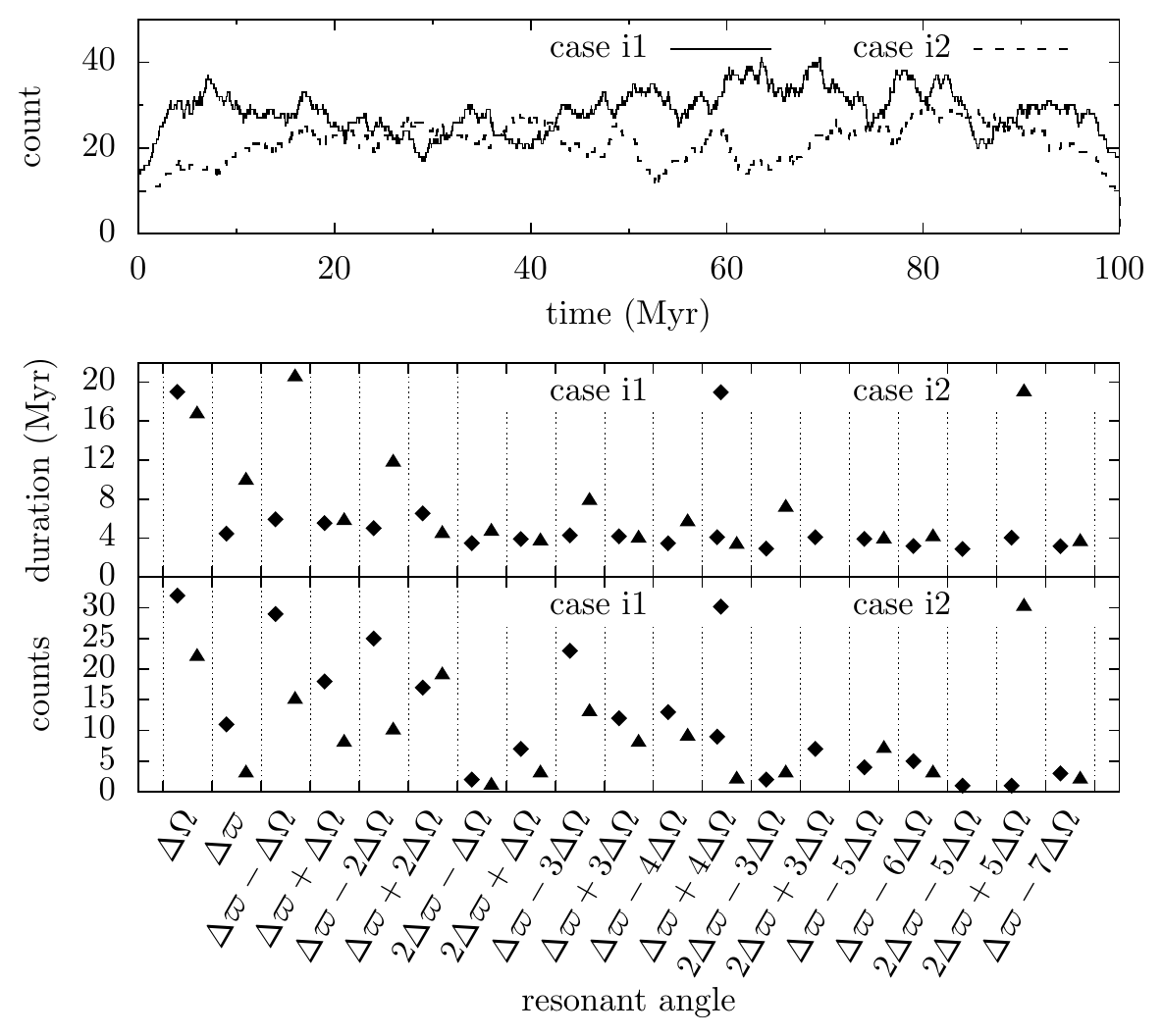}
\caption{The instantaneous number of librating particles, mean resonant passage duration and the total number of particles trapped into each type of resonance for cases i1 (prograde) and i2 (retrograde). Top panel: $x$-axis: time; $y$-axis: number of coexisting librating particles; solid line: case i1; dashed line: case i2. Lower two panels: $x$-axis: the resonant angle; upper $y$-axis: mean resonant duration; lower $y$-axis: number of particles that experience libration at least once; diamonds: case i1; triangles: case i2. Note there are no resonant episodes observed for the angles $2\Delta\varpi+3\Delta\Omega$, $2\Delta\varpi-5\Delta\Omega$ and $2\Delta\varpi+5\Delta\Omega$ in the retrograde case; thus these three columns have no triangular points}
\label{fig-statistics-n}
\end{center}
\end{figure}

The number of librating particles for each type of resonance defines the occurrence rate or frequency. This is shown in the bottom panel of Fig. \ref{fig-statistics-n}. In case i1 (prograde), resonances of $\Delta \Omega$, $\Delta\varpi-\Delta\Omega$, $\Delta\varpi-2\Delta\Omega$ and $\Delta\varpi-3\Delta\Omega$ trapped the most particles ($> 20$), while in case i2 (retrograde), the two resonances with most librating particles (also $\sim20$) involve the angles $\Delta\Omega$ and $\Delta\varpi+2\Delta\Omega$. These low order resonances are relatively strong (Fig. \ref{fig-amp} and Table \ref{tab-region-size}). In both cases, most resonances with $b_1+|b_2|\gtrsim 5$ cannot trap more than $\sim5$ particles. The dependence of the occurrence rate on the coefficients $b_1$ and $b_2$ is different. Resonance trapping occurs for  $b_1$ smaller than 3 but for $b_2$  as large as 7. In addition, we find that the occurrence rate is mainly controlled by $b_1$ -- a resonance with a larger $b_1$ generally traps fewer particles than one with a smaller $b_1$. The explanation may again be related to $e$ and $i$ through Eq. \eqref{eq-order} (Sect. \ref{sec-phoebe-resonance}). In all cases, except $\Delta \varpi +2\Delta \Omega$, $2\Delta \varpi -3\Delta \Omega$ and $\Delta \varpi-5 \Delta \Omega$, prograde resonances trap more particles into libration than their retrograde counterparts.

In total, for the prograde group, there are 221 librating particles and among them, some have experienced more than one type of resonance; thus we have 208 distinct ones. Similarly in the retrograde case, 119 distinct test particles are recorded as 128 librating ones. The fact that the particles can be trapped in more than one resonances (see e.g. the particle in Fig. \ref{fig-retro-n-10-1}) hints that these resonances are close together in $(a,e,i)$ space making resonance overlap possible; cf. Fig. \ref{fig-SEPR}. Specifically for resonances with $b_1+|b_2| \le 5$, 187 distinct prograde and 106 distinct retrograde particles pass through resonances, excluding the apsidal resonance which does not appear in the semianalytical model. These are compared to the relative sizes of the resonant regions in Table \ref{tab-region-size}. We recall that these two particle groups were generated with $\Delta v < 320$ m/s. The results should therefore be compared with the last column, pertaining to the loose group with parameter set i. Good agreement overall is observed. For instance, lower order resonances with larger resonant regions show higher occurrence rates. From Table \ref{tab-region-size}, the total size of all resonant regions with $b_1+|b_2| \le 5$ is $8.4\%$ whereas here the occurrence rates are $18.7\%$ and $10.6\%$ for the prograde and retrograde family respectively.

\begin{figure}[h]
\begin{center}
	\includegraphics[width=0.9\textwidth]{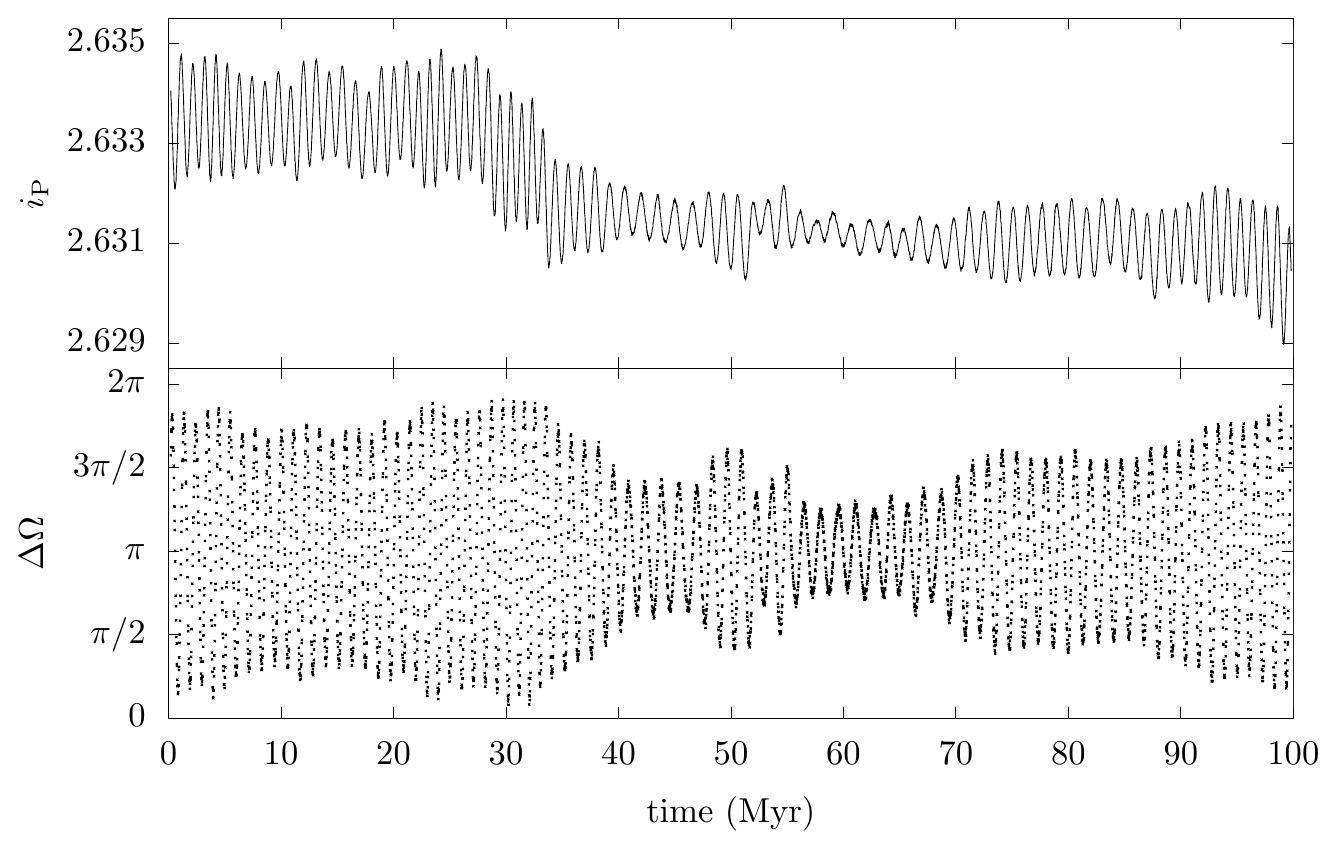}
\caption{A particle in case i2 that persists in nodal libration for the entire integration of 100 Myr. Top panel: inclination; bottom panel: resonant angle $\Delta \Omega$. $x$-axis: time in Myr}
\label{fig-stay-430}
\end{center}
\end{figure}

In Paper I we studied the stability of a given resonance, which we defined as the longest duration of libration episodes of that resonance; there we also presented the distribution of resonant durations. Here, we modify the definition of ``stability'' to be the {\it average} duration of resonant episodes to better represent the most common situations we encounter in our simulations. We calculate the stability of the resonances and present the result in the middle panel of Fig. \ref{fig-statistics-n}. Most resonances cannot trap a particle for more than $\sim 5$ Myr while a few may last for $\sim 20$ Myr. In case i1, the most stable resonance is the nodal resonance whereas in case i2, the resonances of $\Delta \Omega$ and $\Delta \varpi-\Delta \Omega$ are equally stable; all three have $b_1+|b_2|\le 2$. In Paper I, we found two examples of particles remaining in nodal resonance for the entire 100 Myr of the full integration. Here, we observe one such example in case i2, shown in Fig. \ref{fig-stay-430}. Also, for all angles except $\Delta\Omega$, $\Delta\varpi+2\Delta\Omega$ and $\Delta\varpi+4\Delta\Omega$, the retrograde resonances are more stable than their prograde counterparts.

Comparing the occurrence rate and mean resonant duration to the strength in Fig. \ref{fig-amp} and the size of resonant regions in Table \ref{tab-region-size} for case i1, we find that the lower order resonances with larger amplitudes trap more particles and are more stable, as one may intuitively expect. Since we generate the particles randomly in $(a,e,i)$, a resonance with a larger resonant region captures more particles in libration. We may interpret the observed relationship between the mean duration and the resonance strength in the following way: since the particles are slowly diffusing in $(a,e,i)$ as a result of, e.g., close encounters with the massive body, following a random walk \citep[see for instance,][]{Nesvorny2002b,Carruba2003,Delisle2012a}, a larger resonant region may take longer to traverse. This is also observed in case i2, although there are exceptions. For example, capture into the resonance in $\Delta \varpi-\Delta\Omega$, though the most stable in terms of the mean resonant duration, is observed fewer times than a less stable one ($\Delta \varpi+2\Delta\Omega $).

Finally, the top panel of Fig. \ref{fig-statistics-n} shows the number of particles that are librating at a given time. Thus in a loose group around a massive satellite with the mass of Himalia, $\sim 1/30$ of the group members are expected to be in resonance, at any time. This implies that likely none of the four smaller satellites in the actual Himalia group should be currently librating. As shown in \citet{Christou2005}, Lysithea was close to the nodal resonance but probably not in resonance. The findings here seem to support that inference.

\subsection{Statistics of resonances in case ii}
\label{sec-phoebe-statistics}

\begin{figure}[h]
\begin{center}
	\includegraphics[width=0.9\textwidth]{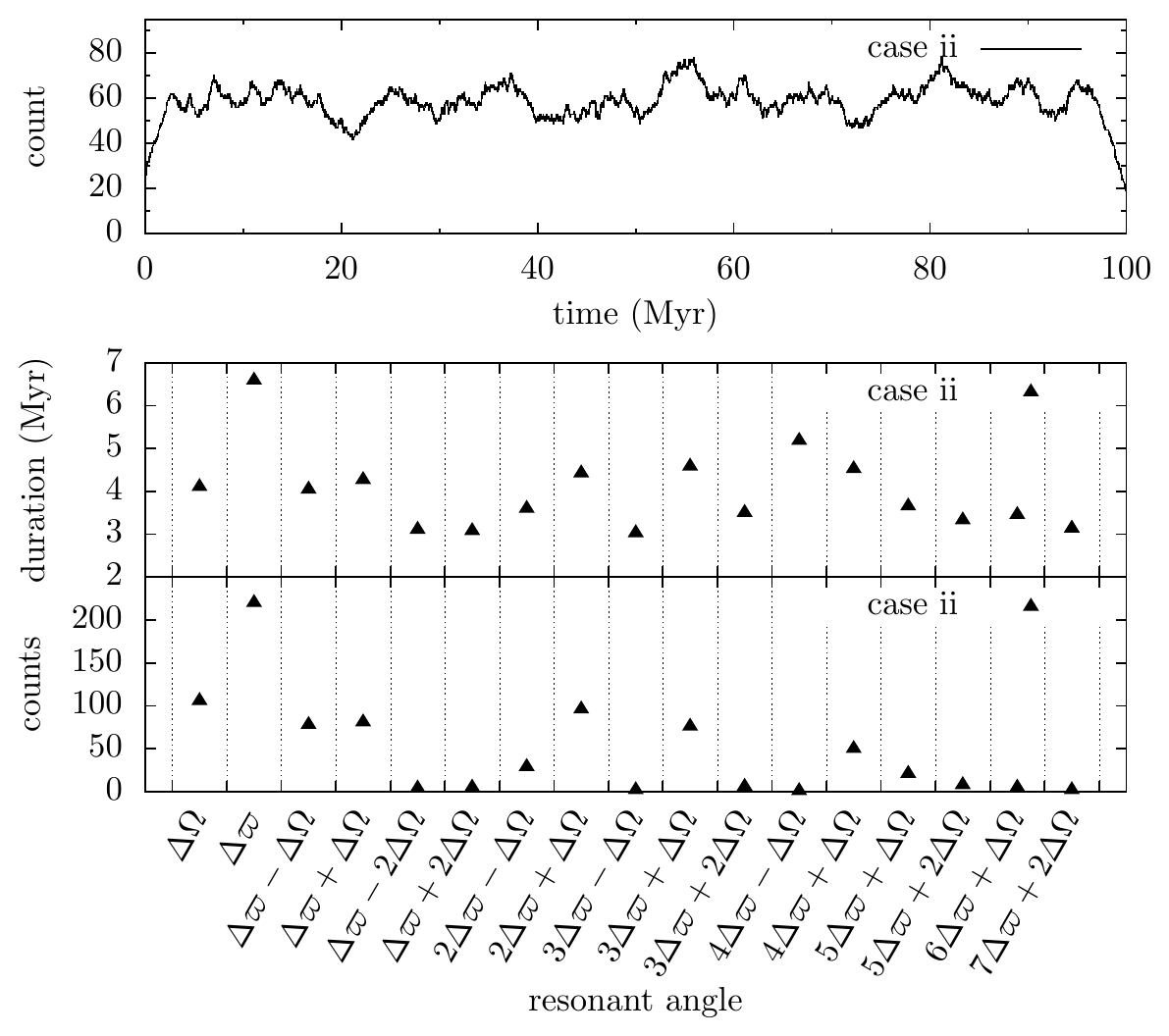}
\caption{As Fig. \ref{fig-statistics-n} but for case ii}
\label{fig-statistics-Phoebe-n}
\end{center}
\end{figure}

An analysis of our simulations of 1000 particles in case ii is presented in Fig. \ref{fig-statistics-Phoebe-n}. Note that these particles are generated under the same  limiting $\delta v$ ($320$ m/s) as for cases i1 and i2. This should be compared with our semianalytical model results for the loose group with parameter set ii, the 4th column of Table \ref{tab-region-size}.

A striking observation is that the apsidal resonance traps the most particles ($>$200), at least twice as many as any other resonance; it also turns out to be the most stable resonance, with mean duration $\approx 6.5$ Myr. Meanwhile, the resonances involving the angles $\Delta\Omega $ and $2\Delta \varpi+\Delta\Omega $ both trap about 100 particles while those for $\Delta\varpi-\Delta\Omega$,$\Delta\varpi+\Delta\Omega$, $3\Delta\varpi+\Delta\Omega$ and $4\Delta\varpi+\Delta\Omega$ all capture between $50 - 80$ particles. These resonances all have $b_2$ being either 0 or 1, which could again be related to the small $i'$ in case ii.

In total, we have detected 790 librating particles among which over a hundred are involved in multiple resonances. So we have 605 distinct librating particles. Out of those, 582 have $b_1+ |b_2| \le 5$ (or 362 if we again exclude the apsidal resonance). The apsidal resonance has the longest mean duration, as discussed above. Curiously, the resonance for $4\Delta \varpi-\Delta\Omega$, though of order 5, is the second most stable. Since this resonance only traps one particle in a single resonant passage, we interpret this as a statistical fluke. Except for the angle $6\Delta \varpi+\Delta\Omega$, all resonances with mean duration $\gtrsim 4$ Myr trap more than 50 librators. So again, those resonances most capable of trapping particles tend to be the most stable.

The semianalytical model predicts that $15.6\%$ of the phase space is influenced by the resonances, in the numerical runs, $36.2\%$ of the particles are captured into at least one type of resonance with $b_1+|b_2|\le 5$. This is similar to what has been observed for cases i1 and i2. We also note a few interesting properties of these resonant episodes that are predicted by the semianalytical model. For example, as shown in the 4th column of Table \ref{tab-region-size}, 6 types of resonances up to order 4 occupy $\ge 0.8 \%$ of phase space; in the numerical simulations, each of them trap $\gtrsim 30$ particles. Table \ref{tab-region-size} also shows that, for the two resonances $b_1 \Delta\varpi- |b_2| \Delta\Omega$ and $b_1 \Delta\varpi+ |b_2| \Delta\Omega$, when $b_1\ge |b_2|$, the latter occupies a larger resonant region than the former-- exactly what we see in the bottom panel of Fig. \ref{fig-statistics-Phoebe-n}.

Apparently, the occurrence rate is generally higher in the numerical simulations than inferred from the size of resonant region in the semianalytical model, especially for case ii (with parameter set ii). In fact, we have been conservative in estimating the resonance width (and thus the resonant region) by using the mean resonant amplitude in the semianalytical model. By definition, it should be the largest amplitude for which the resonance can be maintained \citep[e.g.,][]{Malhotra1996}. Thus the resonant region in the semianalytical model could actually be larger. Furthermore, we have assumed a resonant region to be a sheet of constant thickness, which is not necessarily true; this further complicates the estimate of its size. On the other hand, as mentioned in Sect. \ref{sec-statistics-Himalia}, the particle orbits are slowly diffusing over time, making it possible that a particle would enter a resonant region, though initially outside the resonance. These two factors may have contributed to the difference we observe.

As before, the number of coexisting librating particles is shown in the top panel of Fig. \ref{fig-statistics-Phoebe-n}. On average $\sim 50$ particles are librating simultaneously.

Compared to the retrograde resonances in case i2, here in case ii we observe about four times more resonant particles, probably due to the moon's higher mass and accordingly larger resonant regions. On the other hand, the resonances in case ii seem less stable. This could be caused by the increase in the extent of resonance overlap. Finally, the lower cutoff of the mean duration is $\sim $ 3 Myr. This is an artefact of our method of detecting libration -- only resonant episodes longer than this threshold are accepted as valid. Thus we may have excluded shorter duration resonant episodes from our analysis. In the top panels of Figs. \ref{fig-statistics-n} and \ref{fig-statistics-Phoebe-n} we note that the numbers of concurrent librating particles increase and decrease in the initial and final few Myr of the integration; we attribute this to our adoption of the 3 Myr threshold.

\section{Conclusion and discussion}
\label{sec-conclusion}

Through numerical simulations, we have confirmed that the four resonances between a massive and a massless prograde irregular satellite in nearby orbits as discussed in \citet{Christou2005} and \citet{Li2016} (Paper I) exist for retrograde orbits as well. Considering a restricted four-body problem of planet-Sun-massive satellite-massless satellite (serving as a testparticle), we construct a semianalytical model and find that, after a change of reference system, the retrograde resonances can be described using the same set of equations of motion as for the prograde case for an expansion of the secular interaction potential to lowest order; we show a one-to-one correspondence between critical arguments in the two cases. We use this model to study the evolution of particles on the surface of equal precession rate (SEPR) of an angle of the general form $b_1 \varpi+b_2 \Omega$ and find that all angular differences $b_1 \Delta\varpi+b_2 \Delta\Omega$ with $b_1+|b_2| \le 10$ (except the apsidal resonance where $b_1=1$ and $b_2=0$) may librate, all libration centres being either $0$ or $\pi$ and that the oscillating actions are $e$ and/or $i$ of the particle. For low order arguments ($b_1+|b_2| \le 5$), we present the mean resonant strength and the resonant regions for massive satellites with physical and orbital parameters similar to those of the two irregular satellites, Himalia and Phoebe. We estimate that $\sim20\%$ and $\sim40\%$ of $(a,e,i)$ space in the two cases lies with the resonant regions. We study the survivability of the resonances in the full problem by placing 1000 random particles near satellites with parameters of Himalia and Phoebe; the particles are integrated for 100 Myr with the $N$-body package MERCURY. We find examples of libration for all arguments with $b_1+|b_2|\le 3$ in three cases; some high order resonances also appear. We calculate statistics of the occurrence rate and mean resonant duration, the latter being used as our definition of resonance stability; we observe that a few hundred particles out of the 1000 are trapped in resonances and the typical lifetime of a resonant passage is a few Myr. As intuitively expected, the stronger the resonance, the more particles are temporarily trapped in it and the longer the duration.

As a continuation and extension of Paper I where we concentrated on the four lowest order resonances (up to order 2) here in the current work, we:
\begin{itemize}
\item construct a unified semianalytical model and demonstrate that these resonances induced by a massive body of an orbital group exist for both prograde and retrograde orbits and that there is a one to one connection;
\item generalise the secular resonances to higher order ones of the form $b_1 \Delta\varpi+b_2 \Delta\Omega$ and analyse their likelihood of occurring;
\item show that these secular resonances, even high order ones, can exist for the full Newtonian equations of motion and that the occurrence of these resonances overall agree with prediction.
\end{itemize}

In Paper I, we performed an integration of the observed Himalia group members for 1000 Myr and found that, in addition to Lysithea that was potentially affected by the nodal resonance \citep{Christou2005}, Leda entered the resonance of $\Delta \varpi+\Delta \Omega$. Motivated by our findings here, we revisit that data and find that Leda also experiences a brief passage through the resonance for the angle  $2\Delta \varpi+3\Delta \Omega$.

The secular resonances between close orbits studied here are not restricted to prograde and retrograde irregular satellites. We argue that the existence of such resonances depends on the presence of the surface of equal precession rate (SEPR). As described in detail in Paper I the essence of our approach is that the precession of a small body, say A, is dominated by a strong perturbation (possibly due to another massive body other than the central body); another small body, say B, with a different $a$, $e$ and $i$ can possibly have the same precession rate in an angle -- A and B are on the SEPR of this angle; if B has a small but non-negligible mass, it can change the precession of A slightly (introducing small distortions in the SEPR); A and B are still close to the SEPR; B might have enhanced effects on A due to the commensurability \citep{Greenberg1975}, for instance, causing a secular resonance in the angle difference $b_1 \Delta\varpi+b_2 \Delta\Omega$ and pumping $e$ and $i$. Since the mass of B is small, its influence is limited to its vicinity. One example of such resonances between close orbits involves Ceres \citep{Novakovic2015,Tsirvoulis2016}. Otherwise, if A and B are not typically small, they may mutually interact \citep[probably the close neighbourhood of them has been cleared; see][]{Wisdom1980}. Now the SEPR may be altered by A and B appreciably, introducing further complexity. \citet{Innanen1997} showed the orientations of the orbits of planets in the solar system could be locked with the existence of a hypothetical companion star. \citet{Chen2013} proposed that two planets, under the gravitational effect of an outer residual gas disk, can achieve similar nodal precession rates, causing secular resonances between them.

These secular resonances between close orbits, though excited by a small mass body and probably weak, may have noticeable cumulative effects.  As shown by \citet{Carruba2016a}, the linear secular resonances with Ceres would have depleted the population of nearby asteroids. Similarly, \citet{Christou2005} noticed that the nodal resonance with Himalia may have contributed in the inclination dispersion of that family. These effects are related to the changes in orbital elements after a resonant passage, e.g., change in $i$ after passing through the nodal resonance. In forthcoming work, we intend to study these changes and how they accumulate over time. We also note that for an orbital group around a massive irregular satellite, collisional evolution may be important. Using the method of \citet{Kessler1981} \citep[see also][]{Nesvorny2003} and satellites' radii from \url{http://ssd.jpl.nasa.gov/?sat_phys_par}, we estimate that the expected numbers of collisions in 4 Gyr between a particle and the massive satellite are 0.95 and 0.93 for cases i1 and ii, respectively. Therefore, the effect of collisions should be taken into account when modelling the evolution of actual families.

\begin{acknowledgements}
The authors are grateful for the constructive comments from two anonymous referees, increasing the quality of the paper. We wish to acknowledge the SFI/HEA Irish Centre for High-End Computing (ICHEC) for the provision of computational facilities and support. Astronomical research at the Armagh Observatory is funded by the Northern Ireland Department for Communities (DfC). Fig. \ref{fig-flip-frame} is produced using LibreOffice Draw and Inkscape; all the other figures are generated with gnuplot.
\end{acknowledgements}
\begin{appendices}
\section{Relation of orbital elements in the original and flipped frame}
\label{sec-derivation}
\begin{figure}[h]
\begin{center}
	\includegraphics[width=\textwidth]{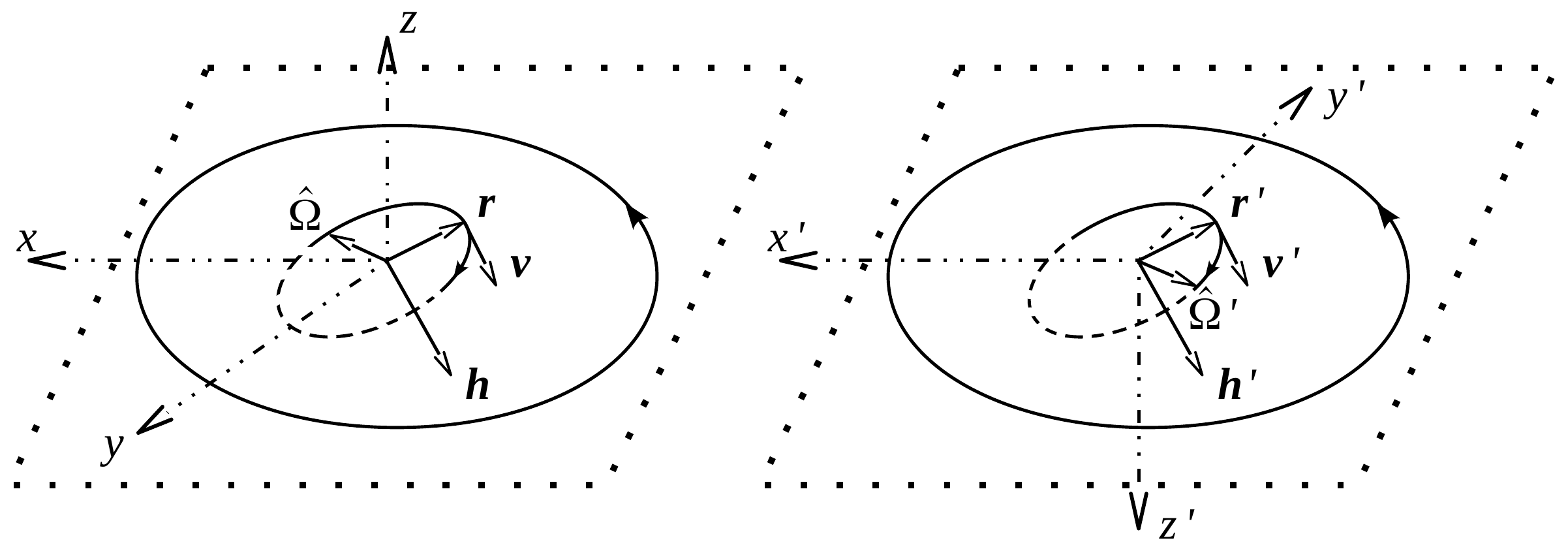}
\caption{The two reference frames: left: the ``original'' fame in which $i_\odot=0$; right: the ``flipped'' frame where $i_\odot'=\pi$ and where we can use the coorbital theory. The planet is at the origin of both frames. The outer dotted parallelogram is the orbital plane of Sun (with respect to the planet) and the big solid ellipse is the orbit of the Sun; the smaller inner inclined one is the orbit of a retrograde satellite; the arrows mark the direction of solar and satellite's orbital revolution. $\vec r$, $\vec v$, $\vec h$ and $\hat \Omega$ are the vectors of position, velocity, angular momentum and the unit vector representing the direction of ascending node in the two frames}
\label{fig-flip-frame}
\end{center}
\end{figure}

We derive the relation of the orbital elements in the two reference systems \eqref{eq-original-flipped-1}. Fig. \ref{fig-flip-frame} is the illustration of the frames and the quantities used in the derivation. As described, each of the frames can be transformed to the other by rotating it along the $x$-axis for $\pi$. We base the derivation on vector transformation in the two frames.

Suppose $\vec P$ is an arbitrary vector and it is $[P_x, P_y, P_z]^\intercal$ in the original frame. Then in the flipped frame, it becomes
\begin{equation}
\label{eq-change-frame}
\vec P'=\left[\begin{array}{c}P'_x\\P'_y\\P'_z\end{array}\right]=\vec R^{x}_\pi \vec P=\left[\begin{array}{ccc}
1 & 0 & 0 \\
0 & -1& 0 \\
0 & 0 & -1 \end{array} \right] \left[\begin{array}{c}P_x\\P_y\\P_z\end{array}\right]=\left[\begin{array}{c}P_x\\-P_y\\-P_z\end{array}\right]
\end{equation}
where $\vec R^{x}_\pi$ is the rotational matrix, meaning rotating the vector along $x$-axis for $\pi$; note ${\vec R^{x}_\pi}^{-1}=\vec R^{x}_\pi$. 

First, we consider the position vector $\vec r$, the velocity vector $\vec v$ and the angular momentum vector $\vec h$. The notations without prime correspond to the original frame and primed ones relate to the flipped frame. Since the transition to the flipped frame is only a rotation, the length of the vectors remain conserved. Thus, eqs. (2.134) and (2.135) of \citet{Murray1999} directly tell $a'=a$ and $e'=e$. According to the definition, the inclination $i'=\arccos (h_z'/\sqrt{h_x'^2+h_y'^2+h_z'^2})=\arccos(-h_z/\sqrt{h_x^2+h_y^2+h_z^2})=\pi-i$. 

The vector representing the direction of ascending node $\hat \Omega$ is frame-specific and we show the calculation of it. Since $\hat \Omega$ is in the orbital plane as well as in the $x-y$ plane. We have $\hat \Omega \perp \vec h$ and $\hat \Omega \perp \hat z$ ($\hat z $ is the unit vertical vector). From the definition of ascending node, we know $\hat \Omega =\hat z \times \vec h=[- h_y, h_x,0]^\intercal$. Thus $\hat \Omega' =\hat z' \times \vec h'=(- h_y', h_x',0)^\intercal=[ h_y, h_x,0]^\intercal$. Since $\Omega$ is measured from $x$-axis, $\sin \Omega=h_x/\sqrt{h_x^2+h_y^2}$ and $\cos \Omega=-h_y/\sqrt{h_x^2+h_y^2}$. Thus
\begin{equation}
\begin{aligned}
\sin \Omega'&=&h_x'/\sqrt{h_x'^2+h_y'^2}&=&h_x/\sqrt{h_x^2+h_y^2}&=&\sin \Omega\,\,
\\
\cos \Omega'&=&-h_y'/\sqrt{h_x'^2+h_y'^2}&=&h_y/\sqrt{h_x^2+h_y^2}&=&-\cos \Omega.
\end{aligned}
\end{equation}
Hence $\Omega'=\pi-\Omega$. Following eq. (2.139) of \citet{Murray1999}, $f$ is a function of $a$, $e$, $h$ (length of $\vec h$), $r$ (length of $\vec r$) and $\dot r$, all remaining the same in the flipped frame, implying $f'=f$. Combining eq. (2.138) with Eq. \eqref{eq-change-frame}, we have $\sin (\omega'+f')=-\sin (\omega+f)$ and $\cos (\omega'+f')=-\cos (\omega+f)$. Hence, $\omega'+f'=\omega+f+\pi$ and $\omega'=\omega+\pi$.

Now we have constructed the relation between the orbital elements in the two frames \eqref{eq-original-flipped-1}.

\end{appendices}

\bibliographystyle{spbasic} 


\end{document}